\documentclass{article}
\usepackage{url}
\usepackage{amsmath}
\usepackage{bm}
\usepackage{dsfont}
\usepackage{amssymb}
\usepackage{multirow}
\usepackage[a4paper]{geometry}
\usepackage[utf8]{inputenc}
\usepackage{lipsum}
\usepackage{natbib}
\usepackage{pbox}
\usepackage{enumitem}
\usepackage{graphicx}
\usepackage{esvect}
\usepackage[export]{adjustbox}
\usepackage[figuresright]{rotating}
\usepackage{natbib}
\usepackage{pdflscape}

\newcommand{\norm}[1]{\left\lVert#1\right\rVert}

\newcommand{\G}{\mathcal{G}}
\DeclareMathOperator{\tr}{tr}
\DeclareMathOperator{\mle}{mle}
\DeclareMathOperator{\gl}{gl}

\DeclareMathOperator{\sgl}{sgl}
\DeclareMathOperator*{\argmin}{arg\,min}

\DeclareMathOperator{\vech}{vech}
\DeclareMathOperator{\vect}{vec}
\DeclareMathOperator{\myvec}{v}

\begin{document}
\title{Fused graphical lasso for brain networks with symmetries}
\date{}
\author{Saverio Ranciati\\
Department of Statistical Sciences, University of Bologna, Italy\\
saverio.ranciati2@unibo.it
\and
Alberto Roverato\\
Department of Statistical Sciences, University of Padova, Italy\\
alberto.roverato@unipd.it
\and
Alessandra Luati\\
Department of Statistical Sciences, University of Bologna, Italy\\
alessandra.luati@unibo.it}

\maketitle

\begin{abstract}
Neuroimaging is the growing area of neuroscience devoted to produce data  with the goal of capturing processes and dynamics of the human brain. We consider the problem of inferring the brain connectivity network from time dependent functional magnetic resonance imaging (fMRI) scans. To this aim we propose the symmetric graphical lasso, a penalized likelihood method with a fused type penalty function that takes into explicit account the natural symmetrical structure of the brain. Symmetric graphical lasso allows one to learn simultaneously both the network structure and a set of symmetries across the two hemispheres. We implement an alternating directions method of multipliers algorithm to solve the corresponding convex optimization problem. Furthermore, we apply our methods to estimate the brain networks of two subjects, one healthy and one affected by mental disorder, and to compare them with respect to their symmetric structure.  The method applies once the temporal dependence characterising fMRI data has been accounted for and we compare the impact on the analysis of different detrending techniques on the estimated brain networks. Although we focus on brain networks, symmetric graphical lasso is a tool which can be more generally applied to learn multiple networks in a context of dependent samples.
\end{abstract}

\section{Introduction}
A brain network is a model of a nervous system represented as a set of nodes, also called vertices, interconnected by a set
of edges; see \citet{bullmore2011brain} for a review on the use of brain graphs for modelling the human brain connectome. Within the domain of human brain mapping, great interest has been posed on the estimation of brain networks from functional magnetic resonance imaging (fMRI) data \citep{smith2011network}.

Functional MRI is a non invasive technique for collecting data on brain activity, with a good resolution in terms of space and time. Essentially, fMRI measures the increase in the oxygenation level at some specific brain region, as long as an increase in blood flow occurs, due to some brain activity. The latent signal in the observed fMRI data is referred to as the blood oxygenation level dependent (BOLD) signal. The BOLD signal arises from the interplay of blood flow, blood volume, and blood oxygenation in response to changes in neural activity.  Under an active state, the local concentration of oxygenated hemoglobyne increases, with a corresponding increase in the homogeneity of magnetic susceptibility, which, in turn, results in an increase of MRI signal. Typically, active states are observed in tasked based experiments in response to an exogenous event.
In recent years, the attention has been concentrating towards Resting state fMRI (RfMRI) data, collected on subjects at rest and in absence of any external stimulus, as the key to understand the neuronal organisation of the brain through the investigation of the spatial and temporal structure of spontaneous neural activity. \citet{Smith13040} carried out an analysis to assess how functional networks at rest match the ones detected under activation tasks. The authors conclude that the resting brain functional dynamics are fully utilising the set of functional networks exhibited by the brain over the range of its possible tasks. In the review paper by \cite{biswal_et_al_2010}, RfMRI is described as the candidate approach capable of addressing the core challenge in neuroimage, i.e. the development of common paradigms for interrogating the  functional systems in the brain, without the constraints of a priori hypotheses.

The construction of a network from fMRI data requires first the identification of a set of functional vertices, such as spatial regions of interest (ROIs), and then the analysis of connectivity patterns across ROIs. It is also relevant that the human brain has a natural symmetric structure. More specifically, it is made up of two hemispheres such that for every spatial ROI on the left hemisphere there is an homologous ROI on the right hemisphere. Accordingly, in the brain network one can identify pairs of homologous vertices and edges, and RfMRI studies have suggested a highly symmetric connectivity; see Section~\ref{sec:data} for additional details.

In this paper, we address the problem of estimating the brain network from RfMRI data by keeping symmetries into explicit account. Special attention is also posed on the temporal dependence characterising fMRI data and the impact of  alternative detrending approaches on the estimated brain network. We remark that we consider undirected graphs, i.e. graphs where the edges are of symmetric type. However, following \citet{hojsgaard2008graphical}, throughout this paper the term ``symmetry'' refers to similarities across the two hemispheres with respect to their network structure and equality in parameter values.

Undirected graphical models \citep{lauritzen1996graphical} are widely applied in network modelling form fMRI data \citep[see, among others,][]{marrelec2006partial,smith2011network,zhu2018sparse}. In this framework, the network structure follows from the sparsity pattern of the inverse covariance matrix of ROI values, and a popular approach to estimate sparse undirected graphical models is the graphical lasso technique \citep{banerjee2008model,friedman2008sparse}.
This method is based on the optimization of a penalized log-likelihood function, where the role of the penalty term is to encourage sparsity in the network. One drawback of the graphical lasso, in the present context, is the fact that it ignores the symmetric structure of the brain. For this reason, we propose a fused graphical lasso method based on the optimization of a penalized log-likelihood where the penalty function is obtained by the sum of two distinct terms. Like the graphical lasso, the first term encourages sparsity in the solution. On the other hand, the second one is a fused type penalty \citep{tibshirani2005sparsity} that encourages symmetry by penalising differences between the left and right hemispheres. As detailed more formally in Section~\ref{sec:method}, symmetries are implemented in the form of equality constraints between entries of the inverse covariance matrix of ROI values. This leads to a convex optimization problem and we provide an alternating direction method of multiplier (ADMM) algorithm for its solution.

The method applies once the temporal dependence characterising fMRI data has been accounted
for and the BOLD signal has been extracted. We assume a simple decomposition
of pre-processed RfMRI data into an unobserved signal plus noise. The underlying hypotheses on
the two latent variables are related to the evolution of the components in time and determine
the method adopted for their estimation. As the dynamics of fMRI time series are controversial, we shall assess the impact of detrending on the estimated ROI connectivity network
using three methods, representative of different approaches to trend estimation, based
on different assumptions and including possible misspecification. In particular, we shall specify a
linear Gaussian multivariate parametric model, a non linear observation driven model for unobserved
components and possibly heavy tailed data, and a non parametric local polynomial
regression method. Details are deferred to Section~\ref{sec:tsa}.

We carry out an extensive analysis and provide an illustration using RfMRI data from two representative subjects who have similar characteristics in terms of age and handedness, though one of the two is healthy while the other has been diagnosed with some mental disorder. We may anticipate that the results
show a lack in the brain asymmetry in the latter individual.

In summary, the novel contribution of this paper is twofold. Firstly, we introduce a fused graphical lasso approach to estimate sparse undirected graphical models with a specific symmetric structure and provide a ADMM algorithm for its solution. An implementation of the latter, written in the R language \citep{RcoreTeam2020}, can be found at \url{https://github.com/savranciati/sgl}. Secondly, we compare the impact on the analysis of different detrending techniques, thereby providing insight into the robustness of the estimated network on such a preliminary step.

\subsection{Related works and possible applications}
The novel contribution of this paper pertains the research area usually referred to as \emph{joint learning of multiple graphical models}. In this framework, the observations come from two or more groups where each group shares the same variables and some of the dependence structure. Accordingly, every group is associated with a network and it is expected that some edges are common across all groups and other edges are unique to each group. More specifically,  the literature has focused on the case where the groups correspond to independent experimental conditions so that every network is a distinct unit, disconnected from the other networks; see, among others, \citet{danaher2014joint,yang2015fused}.
Examples of possible applications include genetic networks \citep[see][]{danaher2014joint} and brain networks from neuroimaging data \citep[see][]{yang2015fused}. In the former, the groups are normal tissue from healthy subjects and one or more different types of cancer tissues whereas in the latter groups correspond to normal subjects and subjects with different degree of cognitive impairment.

Our work is motivated by  brain networks inference where the groups are given by the two hemispheres. The independence assumption does not hold in this case because every observation from the left hemisphere is paired with an observation from the right hemisphere. Thus, existing methods for joint learning of graphical models no longer apply, and one should distinguish between \emph{independent samples} and \emph{paired data}, the latter being a largely unexplored area of research. Note also that, unlike the case of independent samples, with paired data the networks associated with different groups are not expected to be disconnected from each other. Although we focus on brain networks, our approach applies to other settings, and another relevant example comes form cancer genomics where control samples are often obtained from histologically normal tissues adjacent to the tumor (NAT), so that every observation from a cancer tissue is paired with an observation from a normal tissue; see e.g. \citet{aran2017comprehensive}.

A central role in the theory of joint learning of multiple graphical models for independent samples is played by the group and the fused graphical lasso \citep{danaher2014joint}. The penalty term of the group graphical lasso encourages both sparsity and a similar structure of the networks. On the other hand, the fused graphical lasso encourages sparsity and, at the same time, the parameters of the model to be identical across groups. In this way, the groups are encouraged to have both similar network structure and identical parameter values, thereby typically resulting in more parsimonious models. Procedures for applying both the group and the fused graphical lasso are not available for paired data and the symmetric graphical lasso introduced in this paper contributes to fill this gap.

Interestingly, the application of fused graphical lasso for paired data results in a model that belongs to the family of Gaussian graphical models with edge and vertex symmetries, shortly RCON models, introduced by \citet{hojsgaard2008graphical}. Note that, as pointed out in that paper, symmetry restrictions in the multivariate Gaussian distribution have a long history and RCON models can be identified as a special case within this framework. For recent applications of these models see \citet{gao2015estimation,vinciotti2016model,massam2018bayesian}. Although the theory of estimation and testing for  RCON models is well-established, a procedure that performs model selection within the family of RCON models is not available, with the relevant exception of the procedures introduced by \citet{gehrmann2011lattices} and by \citet{li2020bayesian}, within the frequentist  and  the Bayesian approach, respectively, which are of theoretical interest but whose computational complexity restricts their application  to low dimensional settings. More specifically, the problem of model selection for RCON model is discussed in \citet{gehrmann2011lattices} where it is shown that the number of RCON models grows super-exponentially in the number of variables. For this reason, \citet{gehrmann2011lattices} suggested that lasso procedures with fused type penalties might represent a useful alternative to traditional model selection approaches. The symmetric graphical lasso does not constitute a general solution to this problem but it represents, to the best of our knowledge, the first instance of a lasso procedure specifically designed for RCON models.

A further contribution of the paper is concerned with the prewhitening of fMRI data that, measured at each region, are characterised by temporal dependence. There is a longstanding debate on the dynamic properties of fMRI data and parametric models have been employed along with fully non parametric methods. Autoregressive  errors have been considered, see e.g.~\cite{worsley2002general},  \cite{lindquist_2008} and \cite{zhu2018sparse}, as well as fractional noise error processes, as in~\cite{bullmore2003wavelets} or change point methods, see~\cite{aston_kirsch_2012}. Semiparametric methods and high pass filters are also applied to fMRI data, see \cite{Zhang_Yu_2008_AOS} and \cite{schmal2017}, who use the Hodrick-Prescott filter as in \cite{stjohn2015}.  \cite{lund_et_al_2006} deduced that no commonly accepted model for noise in fMRI exists and that regressors may whiten the noise as well as nonparametric smoothing methods. In a Bayesian setting, the relevance of prewhitening has been investigated by \cite{kundu_risk_2020}, who model the temporal covariance  under an inverse-Wishart prior. The  discussion in the latter paper leads to the overall conclusion that prewhitening is a crucial step yet not fully solved when modeling fMRI data.
Against this background, the paper provides a contribution by assessing the impact of different signal extraction methods, applied to the same dataset, to the whitening of the original, temporally correlated, series.


\section{Problem and data description} \label{sec:data}
Structural symmetry of human brain is concerned with anatomical or physiological similarities between the left and the right hemisphere. Otherwise, functional asymmetry is referred to activity-related differences, in a similar way in which left and right hands operate differently, though being anatomically symmetric.
As it is related to behavioral differences, functional asymmetry, also known as lateralisation, is usually detected with respect to some specific tasks, the most relevant being connected to language organisation and handedness. Non invasive methods for exploring the brain organisation with respect to lateralisation are electroencephalography, positron emission tomology, and fMRI, the latter being the most used in research, which  generally display bilateral activations that contrast with the asymmetric effect of lateralisation.

So far, RfRMI studies have suggested a highly symmetric connectivity. Indeed, along with the recognition of the relevance of analysing the brain at rest, the focus has moved from detecting functional asymmetries to detecting structural symmetries. In some sense, the two methods are complementary, but clearly task-based analyses tend to evidence asymmetries whereas resting state analyses are designed to shed light on symmetric structures.  In a recent RfMRI analysis, \cite{Raemaekers2018}  focus on differences between hemispheres that are reflected in asymmetric functional connectivity in resting state subjects and recognise that any asymmetries are prone to be relatively minute. They also observe that a direct quantification of the extent of the hemispheric symmetry is missing.

The fused graphical lasso procedure introduced in this paper provides a methodological contribution for analysing
functional symmetries between the left and right hemisphere of the brain. We apply our  method to a multimodal imaging dataset which comes from a pilot study of the Enhanced Nathan Kline Institute-Rockland Sample project. This project aims at providing a large cross-sectional sample of publicly shared multimodal neuroimaging data and psychological information to 	support and motivate researchers in the relevant scientific goal of understanding the mechanisms underlying the complex brain system. A detailed description of the project, scopes, and technical aspects can be found at \url{http://fcon_1000.projects.nitrc.org/indi/enhanced/}.
The pilot NKI1 study comprises multimodal imaging data and subject-specific covariates for $n = 24$ subjects. Detailed information can be found at \url{http://fcon_1000.projects.nitrc.org/indi/CoRR/html/nki_1.html}.
	
For each subject several information are collected as personal covariates, such as anxiety diagnosis, age, gender, handedness. The fMRI time series are recorded at $p = 70$ spatial Region of Interest (ROI), clustering anatomically close and functionally similar voxels. The way ROIs are defined can depend on the scope of the analysis or on the design of the experiments and it has implications in fMRI analysis, see the discussion by \cite{Poldrack_ROI}. In our case, ROIs are pre-defined according to the Desikan atlas, see \cite{DESIKAN2006968}. For any region, additional information on 3-D spatial locations, hemisphere and lobe membership are available.  As we have ROI-specific information, we apply a region-of-interest  analysis approach, based on the given anatomical parcellation.  An alternative approach is to conduct a whole-brain voxel-wise analysis, on a finer scale, but such an approach is computationally expensive, sensitive to noise, and often difficult to interpret. The optimal means of combining voxels into functionally distinct regions of interest remains to be determined. The issue of parcellation is largely discussed in \cite{craddock2012},  where the authors  develop a spatially constrained spectral clustering approach for group clustering of the whole resting state fMRI data into functionally and spatially coherent regions.
	
As far as dynamic functional activity is concerned, the dataset we are focusing on is composed by time-series data collected for each of the 24 subjects in an imaging session. This imaging technology monitors brain functional activity at different regions via dynamic changes in blood flow creating a low frequency blood oxygen level dependent signal when the subject is not performing an explicit task during the imaging session. In the present NKI1 study, the subjects are simply asked to stay awake with eyes open.
	Focusing on subject $i$ and on scan $k$, where $i=1:24$ and $k=1:2$, we have 70 x 404 matrix whose rows contain the dynamic activity data of the brain regions, collected at T = 404 equally spaced times (time lag is 1400 ms).

The data are provided by Greg Kiar and Eric Bridgeford from NeuroData at Johns Hopkins University, who graciously pre-processed the raw DTI and R-fMRI imaging data available at
\url{http://fcon_1000.projects.nitrc.org/indi/CoRR/html/nki_1.html}, using the pipelines ndmg and C-PAC.

\section{Overview on the methodological framework}\label{sec:method}

Let $X^{(t)}$ be a $p$ dimensional time series vector collecting the fMRI series observed on each single subject over $p=70$ regions,  $t=1,\dots,T$ where $T=404$, the length of each time series. We  assume the general signal plus noise decomposition for $X^{(t)}$,
\begin{equation}\label{eq:spn}
X^{(t)} = M^{(t)} + Y^{(t)}
\end{equation}
where $M^{(t)}$ is the vector collecting the BOLD signal and $Y^{(t)}$ is the idiosyncratic noise component. Our input data for the analysis of the ROI network association structure will be the estimate of $Y^{(t)}$, obtained by contrast as the residual vector once the BOLD signal $M^{(t)}$ is extracted (see Section~\ref{sec:tsa}). More specifically, we denote by $V=\{1,\ldots, p\}$ the set indexing the $p=70$ brain regions
and by $Y_{V}=(Y_{1},\ldots,Y_{70})^{\top}$ the zero mean residual vector, where we have dropped the time index as we assume that the time series dynamics are fully captured by the time varying BOLD signal (Section~\ref{sec:tsa}).

We assume $Y_{V}\sim N_{p}(0, \Sigma)$ and consider the ROI connectivity network obtained from the application of the theory of undirected graphical models \citep{lauritzen1996graphical}. In this framework, the network structure follows from the sparsity pattern of the concentration matrix $\Theta=\Sigma^{-1}$. More specifically, if the entry $\theta_{ij}$ of $\Theta$, with $i\neq j$, is such that $\theta_{ij}\neq 0$ then the brain regions indexed by $i$ and $j$ are connected by an edge in the network. Conversely, for every missing edge in the network the corresponding entry of $\Theta$ is equal to zero. Concentrations can be interpreted by exploiting their connection with partial correlation and regression coefficients, because for every pair $i,j\in V$ with $i\neq j$ it holds that
\citep[see][Section~3.2]{cox1996multivariate},
\begin{align}\label{EQN:interpretation.of.concentrations}
\rho_{ij|V\setminus \{i,j\}}=-\frac{\theta_{ij}}{\sqrt{\theta_{ii}\theta_{jj}}},
\qquad
\beta_{i\leftarrow j|V\setminus \{i,j\}}=-\frac{\theta_{ij}}{\theta_{ii}}
\quad \mbox{and} \quad
\sigma^2_{i|V\setminus \{i\}}=\theta_{ii}^{-1},
\end{align}
where $\rho_{ij|V\setminus \{i,j\}}$ is the partial correlation between $Y_{i}$ and $Y_{j}$ given the remaining components $Y_{V\setminus \{i,j\}}$ whereas $\beta_{i\leftarrow j|V\setminus \{i,j\}}$ is the regression coefficient of $Y_{i}$ on $Y_{j}$ given $Y_{V\setminus \{i,j\}}$ and $\sigma^2_{i|V\setminus \{i\}}$ is the partial variance of
$Y_{i}$ given $Y_{V\setminus \{i\}}$. Hence, if the brain regions indexed by $i$ and $j$ are not connected by an edge it holds that  $\theta_{ij}=0$ and this is equivalent to $\rho_{ij|V\setminus \{i,j\}}=0$ but also to $\beta_{i\leftarrow j|V\setminus \{i,j\}}=0$ and to $\beta_{j\leftarrow i|V\setminus \{i,j\}}=0$. Furthermore, in this case, $Y_{i}$ and $Y_{j}$ are conditionally independent given $Y_{V\setminus \{i,j\}}$.

Every region in the left hemisphere has an homologous region in the right hemisphere so that
the vector $Y_{V}$ can be naturally partitioned into two subvectors. More formally, we set
$q=p/2$ and let the sets $L=\{1,\ldots,q\}$ and $R=\{q+1,\ldots, p\}$ index the subvectors $Y_{L}$ and $Y_{R}$ associated with the left and right hemispheres, respectively, so that the region relative to $Y_{i}$ of $Y_{L}$ is homologous to the region relative to $Y_{i+q}$ of $Y_{R}$; furthermore, to shorten the notation, we set $i^{\prime}=i+q$ for every $i\in L$. Accordingly, the concentration matrix $\Theta$ can be partitioned as
\begin{align*}
\Theta =
\left(
\begin{array}{cc}
\Theta_{LL} & \Theta_{LR}\\
\Theta_{RL} & \Theta_{RR}\\
\end{array}
\right).
\end{align*}
We investigate the presence of symmetries in the ROI association network that take the form of identities between concentrations in  $\Theta_{LL}$ with the corresponding concentrations in $\Theta_{RR}$. This is motivated by the interpretation of such equality restrictions that, by (\ref{EQN:interpretation.of.concentrations}), allows one to identify equality relationships involving partial correlation and regression coefficients. Specifically:
\begin{itemize}
\item[(i)] Equalities involving the diagonal entries imply equality in partial covariances, that is
$\theta_{ii}=\theta_{i^{\prime}i^{\prime}}$ implies $\sigma^2_{i|V\setminus \{i\}}=\sigma^2_{i^{\prime}|V\setminus \{i^{\prime}\}}$.
\item[(ii)] If in addition to the equality
$\theta_{ii}=\theta_{i^{\prime}i^{\prime}}$ in (i) it also holds that  $\theta_{ij}=\theta_{i^{\prime}j^{\prime}}$ then we have  $\beta_{i\leftarrow j|V\setminus \{i,j\}}=\beta_{i^{\prime}\leftarrow j^{\prime}|V\setminus \{i^{\prime},j^{\prime}\}}$ so that the contribution of $Y_{j}$ to the prediction of $Y_{i}$ is identical to the contribution of $Y_{j^{\prime}}$ to the prediction of $Y_{i^{\prime}}$.
\item[(iii)] If in addition to the equalities $\theta_{ii}=\theta_{i^{\prime}i^{\prime}}$ and $\theta_{ij}=\theta_{i^{\prime}j^{\prime}}$ in (ii) it also holds that $\theta_{jj}=\theta_{j^{\prime}j^{\prime}}$ then the partial correlation between $Y_{i}$ and $Y_{j}$ is identical to that  between $Y_{i^{\prime}}$ and $Y_{j^{\prime}}$; formally $\rho_{ij|V\setminus \{i,j\}}=\rho_{i^{\prime}j^{\prime}|V\setminus \{i^{\prime},j^{\prime}\}}$. It is also worth remarking that in this case it follows from (ii) that both $\beta_{i\leftarrow j|V\setminus \{i,j\}}=\beta_{i^{\prime}\leftarrow j^{\prime}|V\setminus \{i^{\prime},j^{\prime}\}}$ and $\beta_{j\leftarrow i|V\setminus \{i,j\}}=\beta_{j^{\prime}\leftarrow i^{\prime}|V\setminus \{i^{\prime},j^{\prime}\}}$.
\end{itemize}

\section{Background}\label{sec:background}
\subsection{Graphical models, graphical lasso and symmetries}
We represent the ROI connectivity network by means of an undirected graph $\G=(V, E)$ where the vertex set $V$ indexes the brain regions and $E\subset V\times V$ is a set of edges, which are unordered pairs of vertices. Let $Y_{V}$ be a  multivariate normal random vector with zero mean vector, variance and covariance matrix $\Sigma=\{\sigma_{ij}\}_{i,j\in V}$ and concentration matrix $\Sigma^{-1}=\Theta=\{\theta_{ij}\}_{i,j\in V}$. The concentration graph model \citep{cox1996multivariate} with graph $\G=(V, E)$ is the family of multivariate normal distributions with $\Theta\in \mathcal{S}^{+}(\G)$, the set of (symmetric) positive definite matrices which have zero elements $\theta_{ij}=0$ whenever $\{i,j\}\not\in E$. The latter model has also been called a covariance selection model \citep{dempster1972covariance} and a graphical Gaussian model \citep{whittaker1990graphical}; we refer the reader to \citet{lauritzen1996graphical} for details and discussion.

Let $S=n^{-1}\sum_{i=1}^{n}y^{(i)}_{V}(y^{(i)}_{V})^{\top}$ be the
matrix of sums of squares and products for a sample $y^{(1)}_{V},\ldots, y^{(n)}_{V}$ of $n$ i.i.d. observations of $Y_{V}$. The maximum likelihood estimator (MLE) $\widehat{\Theta}^{\mle}$
in the concentration graph model with graph $\G$ maximizes the log-likelihood
\begin{align}\label{EQN:log.lik1}
l(\Theta)=\log\det(\Theta)-\tr(S\Theta),
\end{align}
subject to $\Theta\in \mathcal{S}^{+}(\G)$; see \citet[Section~5.2]{lauritzen1996graphical} for details. On the other hand,
the structure of a concentration graph can be estimated from data by determining the zero entries of the concentration matrix. We refer the reader to \citet{drton2017structure} for a comprehensive review on structure learning for graphical models.

In recent years, much interest has focused on the estimation of concentration graph models through the use of $\ell_{1}$ (lasso) regularization. More specifically,  \citet{yuan2007model}, \citet{banerjee2008model} and \citet{friedman2008sparse} proposed the graphical lasso estimator
\begin{align}\label{EQN:glasso.min}
\widehat{\Theta}^{\gl}=\argmin_{\Theta} \left\{-\log\det(\Theta)+\tr(S\Theta)+\lambda |\!|\Theta|\!|_{1}\right\}
\end{align}
where minimization is over the set $\mathcal{S}^{+}$ of $p\times p$ positive definite matrices, $\lambda\geq 0$ and the $\ell_{1}$-norm
$|\!|\Theta|\!|_{1}$ is the sum of the absolute values of the elements of $\Theta$. The graphical lasso adds to the log-likelihood function from (\ref{EQN:log.lik1}) a $\ell_{1}$-penalty pushing the solutions to be sparse, in the sense that
due to the geometry of the $\ell_{1}$-penalty, typically some of the off-diagonal entries of the correlation matrix are shrunk to exactly zero. The term $\lambda$ is the regularization parameter that controls the amount of shrinkage applied to the elements of $\Theta$, and therefore controlling the sparsity of the solution. Thus, graphical lasso is an effective procedure that conducts model selection and estimation simultaneously. Finally, we remark that for $\lambda>0$ the minimum in (\ref{EQN:glasso.min}) is achieved uniquely because the objective is strictly convex, and this holds true also in high-dimensional settings where $p>n$.

\citet{hojsgaard2008graphical} investigated the properties of subfamilies of concentration graph models, named RCON models,
obtained by imposing additional equality restrictions between specified entries of the concentration matrix. RCON models are commonly referred to as colored graphical models because equality constraints can be represented by colouring of edges and vertices of the concentration graph $\G$. Edges of the same color correspond to off-diagonal entries of $\Theta$ with identical values, and similarly for vertices with respect to diagonal entries. The model is thus identified by the structure of $\G$ together with a collection of color classes. \citet{hojsgaard2008graphical} showed that, as well as concentration graph models, RCON models are regular exponential families and provided an algorithm for the computation of the MLE of $\Theta$, implemented in the R package \texttt{gRc} \citep{hojsgaard2007inference}.

\subsection{Time series analysis}\label{sec:tsa}
To assess the impact of detrending on our procedure, we consider three different specifications for the latent components in equation (\ref{eq:spn}). Each one is representative of a wide class of methods for signal extraction and is based on different assumptions on the latent components and their dependence relation. In particular, we specify a Gaussian vector autoregressive model (Section~\ref{subsec:var}), a univariate Student-t score driven model (Section~\ref{subsec:dcs}) and a local polynomial regression filter, that is the Henderson filter (Section~\ref{subsec:lpr}). In the univariate case, we shall denote the elements of the vectors $X^{(t)}, M^{(t)}, Y^{(t)}$ as  $x^{(t)}, \mu^{(t)}, y^{(t)}$, respectively.

\subsubsection{Vector Autoregressive Models}\label{subsec:var}
In the class of multivariate linear models, we consider a first order vector autoregressive process, VAR(1), see \cite{tunnicliffe}, where
$$M^{(t)} = \Phi X^{(t-1)}$$ and $Y^{(t)}$ is assumed to be multivariate normal with zero mean, covariance matrix $\Sigma$ and uncorrelated with $X_{s}$ for $s<t$. The coefficient matrix $\Phi\in \mathbb R^{p\times p}$ is required to have eigenvalues that are in modulus smaller than one and it is usually estimated by least squares. Under distributional assumptions on $Y^{(t)}$ maximum likelihood estimation can be carried out and for VAR processes of higher order, the latter can be selected by means of information criteria.

 \subsubsection{Score driven models}\label{subsec:dcs}
Among nonlinear models for unobserved components, we focus on the class of score driven models,  recently introduced by \cite{creal_2013} and \cite{harvey2013dynamic} as flexible observation driven models for time varying parameters that characterise a given conditional distribution. Specifically, we consider the first order dynamic conditional score (DCS) model for the location
discussed by \citet{harvey2014filtering}, where each $x^{(t)}$ is assumed to be conditionally distributed as a Student-t random variable with $\nu$ degrees of freedom, $x^{(t)}|\mathcal F_{t-1}\sim t_\nu(\mu^{(t)},\sigma^2)$, with the filtration $\mathcal F_{s}$ representing the information set up to time
$s.$ The signal $\mu^{(t)}$ is estimated based on an autoregressive mechanism,
$$\hat\mu^{(t)}   = \omega + \phi \hat\mu^{(t-1)} + \kappa \hat u^{(t-1)}$$ where $\hat u^{(t)}$ is a realisation of a martingale difference sequence, i.e. $E(u^{(t)}|\mathcal F_{t-1}) = 0$, proportional to the score of the conditional likelihood of the time varying location, i.e. $u^{(t)}\propto (\partial  / \partial \mu^{(t)}) \ell(\mu^{(t)}|\mathcal F_{t-1})$, $|\phi|< 1$ and $\hat\mu^{(0)}$ is set equal to a  fixed value.
In this framework, the dynamic BOLD signal is updated by a filter that is robust with respect to extreme values \citep{Calvet_Czellar_Ronchetti2015}. The robustness comes from the properties of the martingale difference sequence $u^{(t)}$: if the data
arise from a heavy tail distribution, then the score $\hat u^{(t)}$ is less sensitive to extreme values than the score of a Gaussian distribution or than the innovation error $\hat v^{(t)} = x^{(t)}-\hat\mu^{(t)}$. An important property of the proposed specification is that it encompasses the Gaussian case, in that the score of the Student-t converges to that of the Gaussian distribution when the degrees of freedom tend to infinity. In practice, if a score driven model is specified when the underlying dataset is in fact Gaussian, a very high value for the degrees of freedom is estimated and a Gaussian model is eventually fitted with the time varying parameter updated through the Kalman filter. The static parameters, $\omega, \nu, \phi, \kappa, \sigma$, are consistently estimated by maximum likelihood and asymptotic standard errors can be derived \cite[see][]{harvey2013dynamic,harvey2014filtering}.

It is important to remark that by applying this method, we are taking into account the possibility that the distribution of the input vector, $Y^{(t)}$, is misspecified, as it is allowed to come from an heavy tailed, rather than Gaussian, distribution.

\subsubsection{Local polynomial regression}\label{subsec:lpr}
Filters that arise from fitting a local polynomial have a well
established tradition in time series analysis and signal
extraction, see \cite{cleveland}.  With no parametric assumptions on the error term,
the signal is approximated locally by a polynomial of degree $d$, so
that in the neighbourhood of time $t$, for $t = h+1,\cdots, n-h$
 one has, for $j = 0,\pm 1, \cdots, \pm h$, $\mu^{(t+j)} = \beta_0+\beta_1 j +\beta_2 j^2 +\cdots+ \beta_d j^d.$
Using this design, the estimate of the trend at time $t$ is simply
given by the intercept, $\hat\mu^{(t)} = \hat\beta_0$. Provided that $2h\geq d$,
the $d+1$ unknown coefficients $\beta_k, k=0,\ldots,d,$ can be
estimated by the method of weighted least squares \citep[see][]{proiettiME} which eventually produce the trend estimate at time $t$ as the result of a weighted average, $$\hat\mu^{(t)} = \sum_{j=-h}^h  w_j x^{(t+j)}.$$

The Henderson filter \citep{henderson1916note} arises as the
weighted least squares estimator of a local cubic trend, i.e. $d=3$, at time
$t$ using $2h+1$ consecutive observations.
Henderson (1916) addressed  the problem of defining a set of
 weights that maximise the smoothness of the estimated
trend, in the sense that the variance of its third differences is minimum. He showed that up to a factor of
proportionality, the resulting weights are the following $w_j \propto
[(h+1)^2-j^2][(h+2)^2-j^2][(h+3)^2-j^2].$

Note that, with local polynomial regression methods, $2h$ trend estimates are missing, corresponding to the first and last $h$ time points. Even if the  latter are not relevant in the present paper, the reader is referred to \citet{proiettiAOAS} for estimation of the signal at the boundaries by asymmetric filters.

\section{The symmetric graphical lasso}\label{sec:fusedglasso}

\subsection{The penalized log-likelihood}
In order to encourage both sparsity in the graph structure and similarity across the two brain hemispheres, we introduce a specific fused-type penalty \citep{hoefling2010path,tibshirani2011solution} especially designed to encourage the equality between the concentration values of the relevant subgraphs.  Hence, we propose the following estimator of $\Theta$, which we name the symmetric graphical lasso estimator,
\begin{align}\label{lik2}
\widehat{\Theta}^{\sgl} =
\argmin_{\Theta}\{-\log\det(\Theta)+\text{tr}(S\Theta)
+\lambda_1\!\norm{\Theta}_1
+\lambda_2\!\norm{\Theta_{LL}-\Theta_{RR}}_1\},
\end{align}
where $\lambda_{1},\lambda_{2}\geq 0$ are regularization parameters that control the
amount of shrinkage. Equation (\ref{lik2}) is obtained by adding to the (minus) log-likelihood in (\ref{EQN:log.lik1}) a convex penalty function obtained as the sum of two $\ell_{1}$-penalties, i.e. the penalty  $\lambda_{1}|\!|\Theta|\!|_{1}$ that, like the graphical lasso, for large values of $\lambda_{1}$ encourages sparsity in $\widehat{\Theta}^{\sgl}$, and the penalty $\lambda_2\!\norm{\Theta_{LL}-\Theta_{RR}}_1$
that, for large values of $\lambda_{2}$ encourages the elements of $\widehat{\Theta}^{\sgl}_{LL}$ to be identical to the corresponding elements of $\widehat{\Theta}^{\sgl}_{RR}$ \citep{tibshirani2005sparsity,danaher2014joint}. Recall that, as described in Section~\ref{sec:method}, such equality constraints may, in turn, imply the equality of other quantities of interest, such as regression coefficients and partial correlation coefficients.

One of the appealing features of the lasso is that it typically performs model selection and estimation simultaneously. From this perspective, it is worth remarking that the symmetric graphical lasso performs model selection and estimation within the class of RCON models. More precisely, it is suited to identify color classes of the form $\{\theta_{ij}, \theta_{i^{\prime}j^{\prime}}\}$ corresponding to  $\hat{\theta}^{\sgl}_{ij}=\hat{\theta}^{\sgl}_{i^{\prime}j^{\prime}}$, which are of natural interest in the analysis of brain networks.

\subsection{An algorithm for the symmetric graphical lasso problem}\label{subsec:nestedADMM}
In order to solve equation (\ref{lik2}) we use an alternating direction method of multiplier (ADMM) algorithm. A comprehensive exposition of ADMM algorithm can be found in \citet{boyd2011distributed} whereas we refer to \citet{danaher2014joint} and \citet{tan2014learning}, and references therein, for applications of ADMM to related problems. ADMM is an attractive algorithm for this problem because it allows us to split the optimization procedure into two nested, less involved, convex optimization problems. These can be both solved using suitable ADMM algorithms.

First, we note that the optimization problem in (\ref{lik2}) is equivalent to  minimize with respect to $\Theta$ and $Z$ the quantity
\begin{equation}\label{min1}
-\log\det(\Theta)+\tr(S\Theta)+\lambda_1\!\norm{Z}_1+\lambda_2\!\norm{Z_{LL}-Z_{RR}}_1,
\end{equation}
where $\Theta$ and $Z$ are restricted to belong to $\mathcal{S}^{+}$ and subject to the linear constraint $Z=\Theta$. We remark that $Z_{LL}$ and $Z_{RR}$ in (\ref{min1}) are the relevant diagonal submatrices of $Z$. Hence, the scaled form of the augmented Lagrangian can be written as \citep[Section~3.1.1]{boyd2011distributed},
\begin{align}\label{lagr1} \text{L}_{\rho_{1}}\bigl(\Theta,Z,U\bigl)
&=-\log\det(\Theta)+\tr(S\Theta)+\lambda_1\!\norm{Z}_1+\lambda_2\!\norm{Z_{LL}-Z_{RR}}_1+ \nonumber \\
&+\dfrac{\rho_1}{2}\norm{\Theta-Z+U}^2_{\text{F}} - \dfrac{\rho_1}{2}\norm{U}^2_{\text{F}},
\end{align}
where $U$ is the scaled dual variable and the symbol $\norm{\cdot}_{\text{F}}$ denotes the Frobenius norm, i.e. the square root of the sum of the squared entries  of its argument. The ADMM algorithm for the optimization of (\ref{lagr1}) uses the augmented Lagrangian parameter  $\rho_1>0$ as `step size' and, when the algorithm is at convergence, due to the constraint $Z=\Theta$, the last two terms of equation (\ref{lagr1}) cancel out and one obtains the solution to (\ref{lik2}).

ADMM iterates three fundamental steps in order to minimize (\ref{lagr1}) \citep[see][equations (3.5) to (3.6)]{boyd2011distributed}. More formally, we initialize $Z^{1}$ and $U^{1}$ equal to the zero matrix and for $l=1,2,3,\ldots$
the updates for the quantities $(\Theta, Z, U)$ are obtained as \citep[see also][Section~6.6]{boyd2011distributed}:
\begin{enumerate}[label=(\arabic*)]
\item $\displaystyle \Theta^{l+1}:=\argmin_{\Theta}\left( -\log\det(\Theta)+\tr(S\Theta)+\dfrac{\rho_1}{2}\norm{\Theta-Z^l+U^l}^2_{\text{F}} \right);$
\item $\displaystyle Z^{l+1}:=\underset{Z}{\argmin}\biggl(\lambda_1\!\norm{Z}_1+\lambda_2\!\norm{Z_{LL}-Z_{RR}}_1+\dfrac{\rho_1}{2}\norm{\Theta^{l+1}-Z+U^l}^2_{\text{F}}\biggl);$
\item $\displaystyle U^{l+1}:=U^l+\Theta^{l+1}-Z^{l+1}.$
\end{enumerate}

The implementation of step (3) is straightforward and in the following we describe steps (1) and (2) in detail.

Step (1) has an analytical solution, with computational complexity given by performing an eigendecomposition of a $p \times p$ matrix. More specifically, if $QDQ^{\top}$ is the eigendecomposition of $\rho_{1}(Z^{l}-U^{l})-S$ then the solution is given by $\Theta^{l+1}:=Q\tilde{D}Q^{\top}$ where $\tilde{D}$ is the diagonal matrix with $i$th diagonal entry $(d_{ii}+\sqrt{d_{ii}^{2}+4\rho_{1}})/(2\rho_{1})$ and $d_{ii}$ is the $i$th diagonal entry of $D$. Note that the diagonal entries of $\tilde{D}$ are always positive because $\rho_{1}>0$ and therefore $\Theta^{l+1}\in\mathcal{S}^{+}$, as required. Finally, we remark that this step of ADMM coincides with the corresponding step of ADMM for graphical lasso and the reader can see \citet[Section~6.6]{boyd2011distributed} for further details.

We turn now to step (2) of the algorithm. For a matrix $Q$ with rows and columns indexed by $V=L\cup R$  we let $\myvec(Q)$ be the vector defined as
\begin{align*}
\myvec(Q)^{\top}=
\left[
\begin{array}{ccc}
\vech(Q_{LL})^{\top} &
\vech(Q_{RR})^{\top} &
\vect(Q_{LR})^{\top}
\end{array}
\right],
\end{align*}
where $\vect(\cdot)$ and $\vech(\cdot)$ are the vectorization and half-vectorization operators, respectively. Hence, we set
\begin{align*}
z = \myvec(Z),\quad b^{l} = \myvec(\Theta^{l}) + \myvec(U^{l})\quad\mbox{and}\quad
F=\left[
\begin{array}{ccc}
 I & -I & O
\end{array}
\right]
\end{align*}
where $I$ is the identity matrix of dimension $q(q+1)/2$ and $O$ is the $q(q+1)/2\times q^{2}$ zero matrix. We can thus write the second step of the main ADMM algorithm in the form,
\begin{equation}\label{genlas}
\argmin_{z} \biggl(\dfrac{1}{2}\norm{z-b}^2_{2}+\lambda^{\prime}_{1}\!\norm{z}_1+\lambda^{\prime}_{2}\norm{Fz}_1\biggl),
\end{equation}
where  $\norm{\cdot}_2$ is the Euclidean norm, $\lambda^{\prime}_1=\lambda_{1}/\rho_{1}$, $\lambda^{\prime}_{2}=\lambda_{2}/\rho_{1}$ and, to simplify the notation, we have omitted the superscript from $b$. Equation (\ref{genlas}) shows that the optimization problem in the second step of ADMM is a special variant of the classical fused lasso called the fused lasso signal approximator. This allows us to exploit known results for this class of problems. More specifically, \citet[Lemma A.1]{friedman2007pathwise} showed that if a solution of (\ref{genlas}) for $\lambda^{\prime}_{1}=0$ and $\lambda^{\prime}_{2}>0$ is known, then the solution for $\lambda^{\prime}_{1}>0$ can be easily obtain in closed form through a soft-thresholding operation. Hence, we can focus on the solution of
\begin{equation}\label{genlas.reduced}
\argmin_{z} \biggl(\dfrac{1}{2}\norm{z-b}^2_{2}+\lambda^{\prime}_{2}\norm{Fz}_1\biggl),
\end{equation}
that is a generalized lasso problem \citep{tibshirani2011solution}, and
an ADMM algorithm for its solution can be found in  \citet[Section~6.4.1]{boyd2011distributed}. Concretely, the ADMM algorithm iterates until convergence through the following steps:
\begin{enumerate}[label=(\roman*)]
\item $z^{m+1}:= \bigl(I+\rho_2F^{\top}\!F\bigl)^{-1}\left\{b+\rho_2F^{\top}\!(v^m-t^m)\right\}$;
\item $v^{m+1}:=\mathcal{S}_{\lambda'_2/\rho_2}(Fz^{m+1}+t^m)$;
\item $t^{m+1}:=t^m+Fw^{m+1}-v^{m+1}$
\end{enumerate}
In step (i), $I$ is an identity matrix of appropriate dimension and $\rho_2>0$ is the `step size' for the inner ADMM. In step (ii), $\mathcal{S}_{\kappa}(\cdot)$ is the soft thresholding operator \citep[see][Section~4.4.3]{boyd2011distributed}. The vectors $v$ and $t$ mimic the role of $Z$ and $U$ of the outer ADMM and can be initialized to the zero vector. If we denote by $z_{[\lambda^{\prime}_{1}=0, \lambda^{\prime}_{2}]}$ the optimal solution at convergence of (\ref{genlas.reduced}), then we can apply \citet[Lemma A.1]{friedman2007pathwise} and adjust $z_{[\lambda^{\prime}_{1}=0, \lambda^{\prime}_{2}]}$ element-wise to obtain the optimal solution of (\ref{genlas}) for a given $\lambda^{\prime}_{1}>0$ as
$z_{[\lambda^{\prime}_{1}, \lambda^{\prime}_{2}]}=\mathcal{S}_{\lambda_1/\rho_1}(z_{[\lambda^{\prime}_{1}=0, \lambda^{\prime}_{2}]}).$
The update $Z^{l+1}$ for step (2) of the outer ADMM algorithm is thus the symmetric matrix such that
$\myvec(Z^{l+1})=z_{[\lambda^{\prime}_{1}, \lambda^{\prime}_{2}]}$.

Finally, as stopping rule we use a tolerance check on the total relative change of the current estimate of the solution. In particular, if $\frac{|| \Theta^{m}-\Theta^{m-1}||}{||\Theta^{m-1}||}$ is lower than the chosen tolerance, the algorithm is stopped and assumed to be at convergence.

\section{Simulation study}\label{sec:simulations}
We carry out a simulation study which aims to assess the performance of symmetric graphical lasso in a framework that mimics the structure of the RfMRI data in Section~\ref{sec:data}. For this reason, we apply our procedure to simulated datasets sampled  from normally distributed random vectors $Y_{V}$ of $|V|=p=70$ variables with $V=L\cup R$, as in  Section~\ref{sec:method}. We consider two scenarios, denoted by A and B, that differ in their edge and symmetry degrees, with scenario A being sparser than B. The experiment is designed as follows. First, we randomly generate two undirected graphs, $\G^{A}$ and $\G^{B}$, with edge degrees, computed as the ratio of the number of edges of the graph over the number of edges of the complete graph, $p(p-1)/2$, equal to $d^{A}=23.1\%$ and $d^{B}=31.6\%$, respectively. Next, for each scenario, we randomly generate 4 positive definite concentration matrices $\Theta^{A}_{\text{true},i}$ and $\Theta^{B}_{\text{true},i}$, for $i=1,\ldots,4$ with zero pattern corresponding to the missing edges of $\G^{A}$ and $\G^{B}$, respectively. The concentration matrices are constructed in order for a given proportion of randomly selected pairs of homologous concentrations across the two brain hemispheres to have the same value. More specifically, we focus on present edges, with nonzero concentration values, and the proportion of pairs of symmetric nonzero concentration is $d^{A}_{\text{sym}}=10.8\%$ for scenario A and $d^{B}_{\text{sym}}=30.1\%$ for scenario B.
The 8 generated concentration matrices characterize 8 normal distributions with zero mean vector, and from each of these distributions we extract 9 i.i.d. samples of size $n=400$, so as to resemble the sample size of the data in Section~\ref{sec:data}.

In the penalized likelihood framework, a controversial question is how to choose the regularization parameter and several methods have been proposed in the literature. This issue is beyond the scope of this paper and, in order to avoid that the performance of symmetric graphical lasso is confounded by the choice of a selection method, we follow an ``oracle'' procedure, described in the following.
In each of the 72 generated datasets, we apply the graphical lasso and choose the value of $\lambda_1$ that produces a graph with an edge density equal to the density of the graph used to simulate the data. Next, conditional on the selected value of $\lambda_1$, we apply the symmetric graphical lasso for $10$ different logarithmically spaced values of $\lambda_2$.

For every selected model, we consider some well established measures to assess the performance in  recovering the graph structure. The same quantities are then adapted to assess the performance in  recovering the symmetric structure. Specifically, we compute the edge positive-predicted value (ePPV), also called precision, as the ratio between the number of true edges (eTP) and the number of edges (\#edges) in the selected graph, and the symmetry positive-predicted value (sPPV) as the ratio between the number of true symmetric nonzero concentrations (sTP)  and the number of nonzero symmetric concentrations (\#symm) in the estimated concentration matrix. Furthermore, we compute the edge true-positive rate (eTPR) as the ratio between eTP and the number of edges (eP) in the true graph and the symmetry true-positive rate (sTPR) as the ratio between sTP and the number symmetric nonzero concentrations (sP) in the true concentration matrix. Similarly, we compute the edge true-negative rate (eTNR) and the symmetry true-negative rate (sTNR). In this way, we consider the quantities

\begin{align}\label{eqn:four.performance.measures-E}
\mbox{ePPV}=\frac{\mbox{eTP}}{\mbox{\#edges}},
\qquad
\mbox{eTPR}=\frac{\mbox{eTP}}{\mbox{eP}}
\quad
\mbox{and}
\quad
\mbox{eTNR}=\frac{\mbox{eTN}}{\mbox{eN}},
\end{align}
which we use to asses how much the symmetric graphical lasso procedure recovers the graph structure, and the quantities
\begin{align}\label{eqn:four.performance.measures-S}
\mbox{sPPV}=\frac{\mbox{sTP}}{\mbox{\#symm}},
\qquad
\mbox{sTPR}=\frac{\mbox{sTP}}{\mbox{sP}}
\quad
\mbox{and}
\quad
\mbox{sTNR}=\frac{\mbox{sTN}}{\mbox{sN}},
\end{align}
used to asses the ability of the symmetric graphical lasso to identify symmetries. It is worth remarking that symmetric graphical lasso tends to encourage equality between both zero and nonzero concentrations but we evaluate its performance only with respect to nonzero concentrations whose identification is of greater interest in applied contexts.

Table~\ref{tab_sim1} summarises the behaviour of the symmetric graphical lasso for a dataset in the scenario A. More specifically, we report the performance measures for the model selected by the graphical lasso and each of the 10 models obtained from the 10 values of $\lambda_{2}$ in the application of the symmetric graphical lasso.  As shown in the first two lines of Table~\ref{tab_sim1}, the results for the graphical lasso and the symmetric graphical lasso with the lowest value of $\lambda_2$ are virtually identical, which is expected given the equivalence between our proposed approach and graphical lasso when $\lambda_2=0.$ Increasing values of $\lambda_2$ tend to increase the sparsity of the selected graph,
which can be explained
by the fact that symmetric graphical lasso encourages symmetries also for zero concentrations. As a consequence, increasing values of $\lambda_{2}$ tend to correspond to a moderate decrease in the values of ePPV and eTPR; on the other hand, eTNT tends to increase as $\lambda_2$ increases.

To choose the value of $\lambda_{2}$ we adopt the ``oracle''  criterium that selects the model corresponding to
the highest sum sTPR$+$sTNR, highlighted in bold in Table~\ref{tab_sim1}. We note that this corresponds to the sparsest graph, with $496$ present edges.  Focusing  on the symmetric concentrations, we see, as expected, an increase in sTPR for increasing values of $\lambda_2$, with a steady decrease in terms of sTNR, with an appreciable value of 89.92 for the selected model. If we piece together both the considerations, we can see from Table~\ref{tab_sim1} that the price in terms of eTPR and eTNR paid by using symmetric graphical lasso instead of the graphical lasso is worth the additional information we gain by recovering the symmetric structure of the two blocks of the concentration matrix, and the associated graph.

\begin{figure}
\centering
\includegraphics[width=0.72\textwidth]{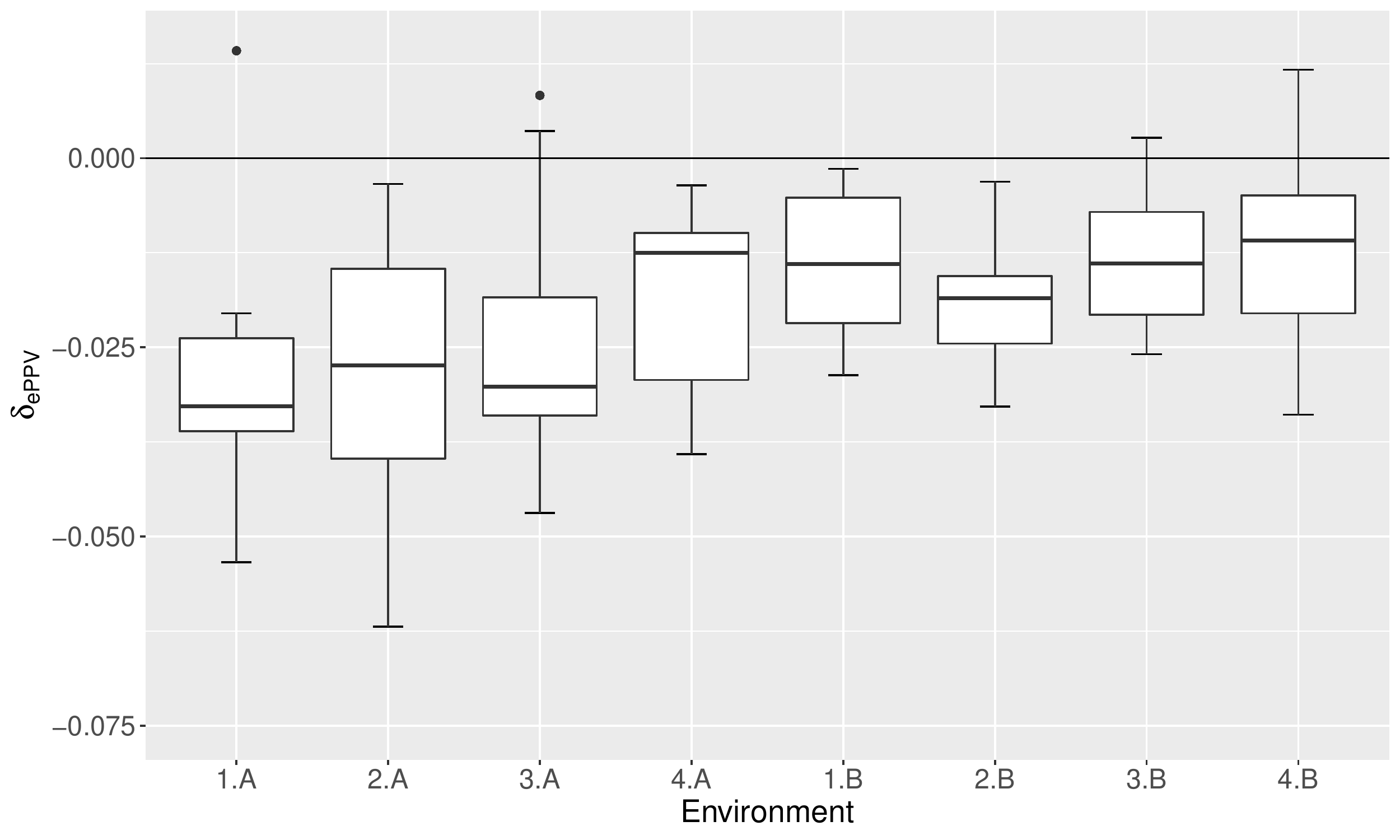}
\\
\includegraphics[width=0.72\textwidth]{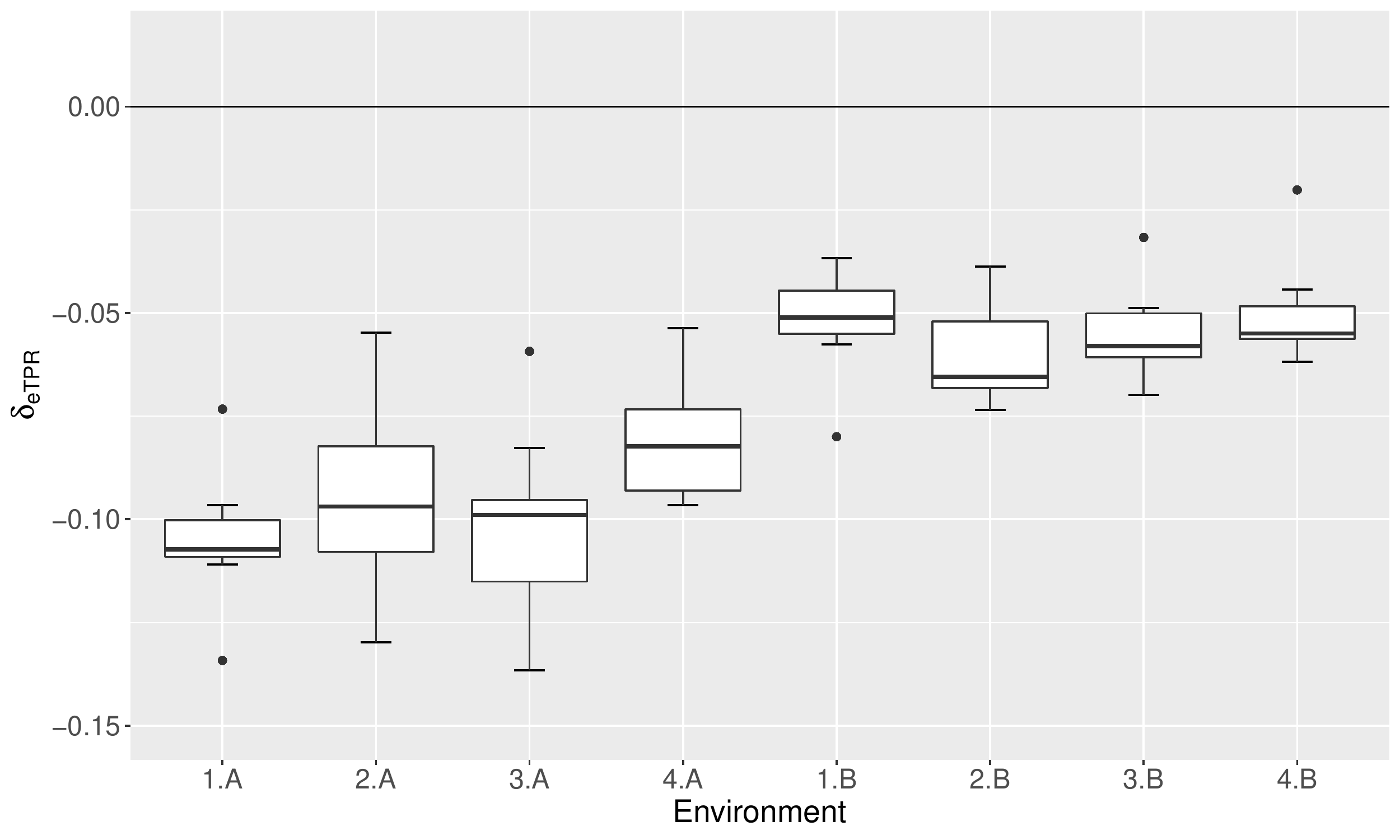}
\\
\includegraphics[width=0.72\textwidth]{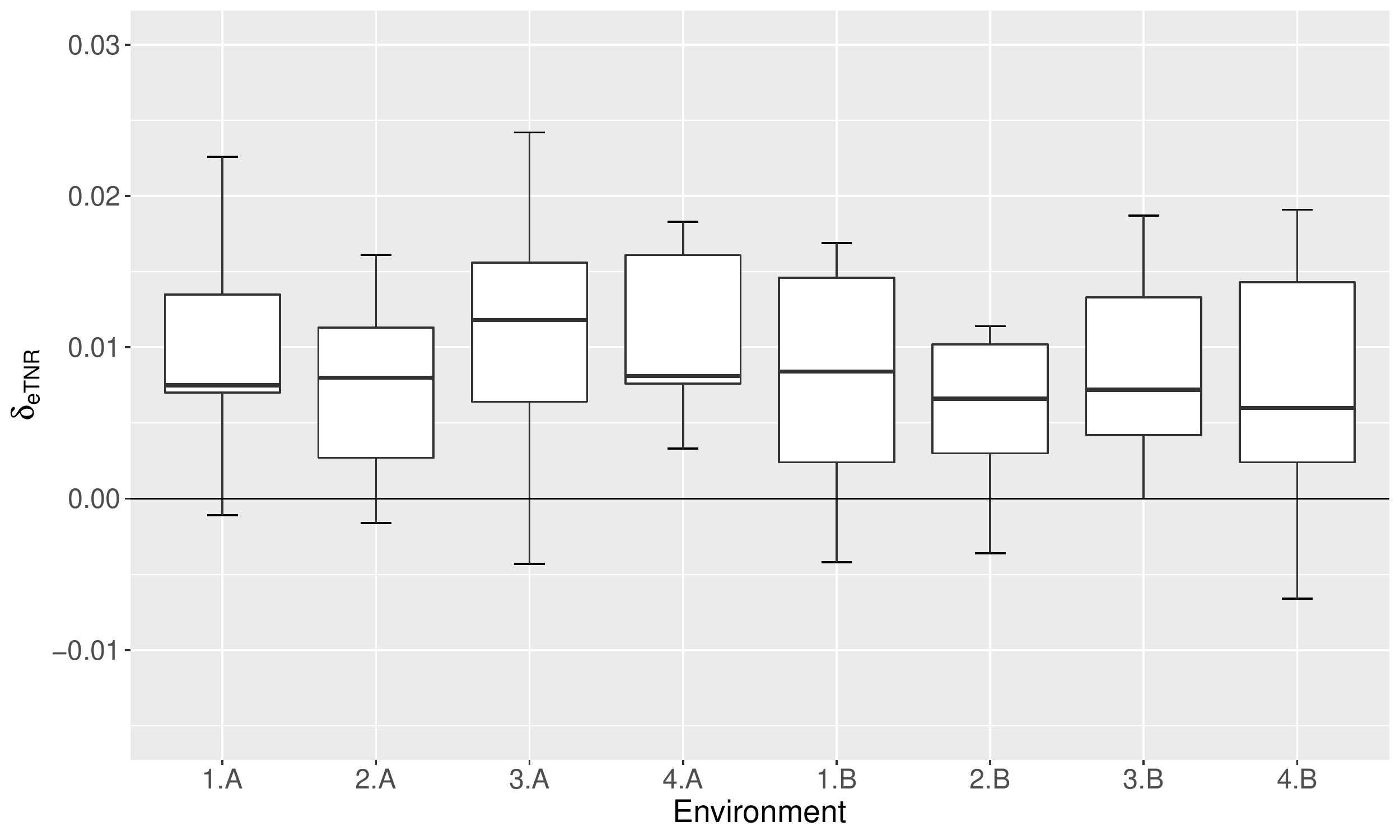}
\caption{Difference between a given performance measure (top panel: ePPV;  middle panel: eTPR; bottom panel: eTNR) computed on the model selected by symmetric graphical lasso and the same measure computed on the model selected by graphical lasso, for each of the $8\times 9$ datasets. Every boxplot summarises the 9 datasets of the corresponding environment.}
\label{fig_sim1}
\end{figure}

We apply this procedure to the 72 generated datasets thereby obtaining 72 models identified by graphical lasso and 72 models identified by symmetric graphical lasso. The result of these analyses are summarized in Table~\ref{tab_sim2} and Figure~\ref{fig_sim1}. These show that, as far as the structure of the graph is concerned, the graphs obtained from symmetric graphical lasso have smaller values of eTPR and ePPV with respect to graphical lasso, and higher values of eTNR. However, the reduction in eTPR and ePPV is moderate in the sparser scenario A, about 10\% for eTPR and 2.5\% for ePPV, and, in fact, quite small in  the denser scenario B, about 5\% for eTPR and 1.5\% for ePPV.
Nonetheless, these moderate reductions in eTPR and ePPV are compensated by an increase in eTNR and, most importantly, by a satisfying performance in terms of recovery of symmetries. Interestingly, the symmetric graphical lasso seem to have a very similar behaviour in the two scenarios as far as sTPR and sTNR are concerned, but scenario B shows higher values of sPPV.

\section{Analysis of RfMRI data}\label{sec:application}
The symmetric graphical lasso is applied to the data described in Section~\ref{sec:data}. The analysis is focused on two subjects, indexed as subject 18 and subject 22, who show homogeneous characteristics in terms of age and handednenss but different diagnosis status. According to the available explanatory variables, subject 18 is 46 years old, right-handed, and healthy, whereas subject 22 is 42 years old, right-handed as well, but had a current and recurring diagnosis of drug abuse and mental disorders at the time of the fMRI scan recording.

We first account for the temporal dependence, with the tools and methodologies discussed in Section~\ref{sec:tsa}. In particular, for each subject, we obtain a matrix of residuals of dimension $n \times p$ from a VAR(1) model, a first order score driven model, and a Henderson filter with  $h=6$ (a 13-term weighted average). We then apply the symmetric graphical lasso to the residuals. As criteria for choosing the optimal value of $\lambda_{1}$ and $\lambda_{2}$ we use the Bayesian Information Criterion (BIC) and the extended BIC  \citep[eBIC;][]{foygel2010extended},
\begin{align*}
\text{eBIC}= -2l(\widehat{\Theta}^{\mle})+\log(n)\,d + 4 d \gamma \log(p),
\end{align*}
where $l(\widehat{\Theta}^{\mle})$ and $d$ denote the maximized log-likelihood function and the number of free parameters of the relevant model, respectively. The eBIC depends on the parameter $\gamma\in [0;\,1]$ that controls how much the criterion prefers simpler models. The limit case  $\gamma=0$ corresponds to the classical BIC. As suggested in \citet{foygel2010extended} we set $\gamma=0.5$. For the computation of the maximum likelihood estimates within the family of RCON models, we use the \texttt{gRc}  package for R \citep{hojsgaard2007inference}. As a joint grid search over $\lambda_1$ and $\lambda_2$ could be computationally prohibitive \citep[see also][]{danaher2014joint}, we fix first $\lambda_2$ to a low value - close to zero - while performing a dense grid search over $\lambda_1.$ After selecting the best value of $\lambda^{\star}_1$, a conditional sweep on a grid of 20 equally spaced values on a logarithmic scale for $\lambda_2$ can be performed to select the final pair of optimal values $(\lambda^{\star}_1,\lambda^{\star}_2).$

\begin{landscape}
\begin{figure}
\centering
\includegraphics[width=0.46\textwidth]{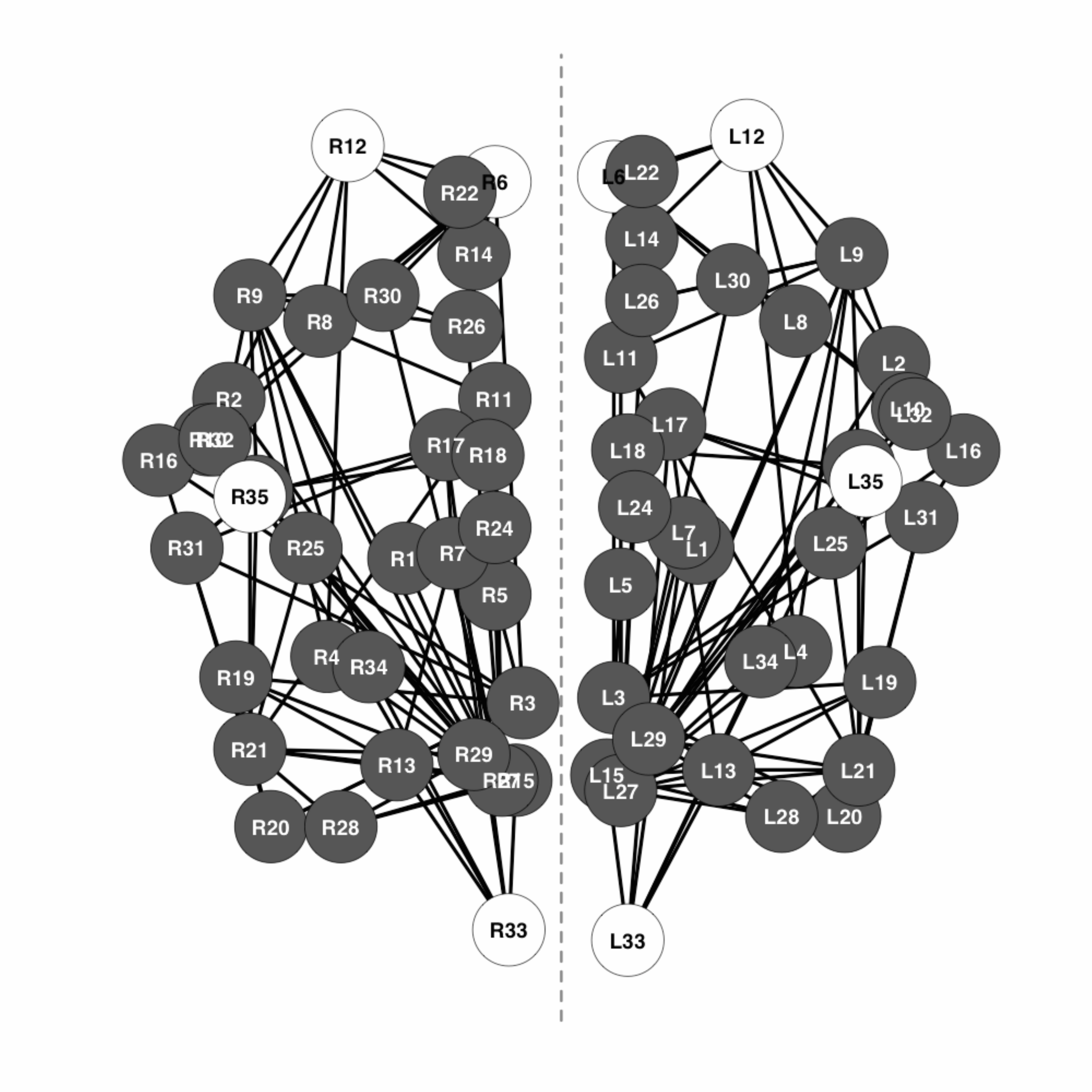}\includegraphics[width=0.46\textwidth]{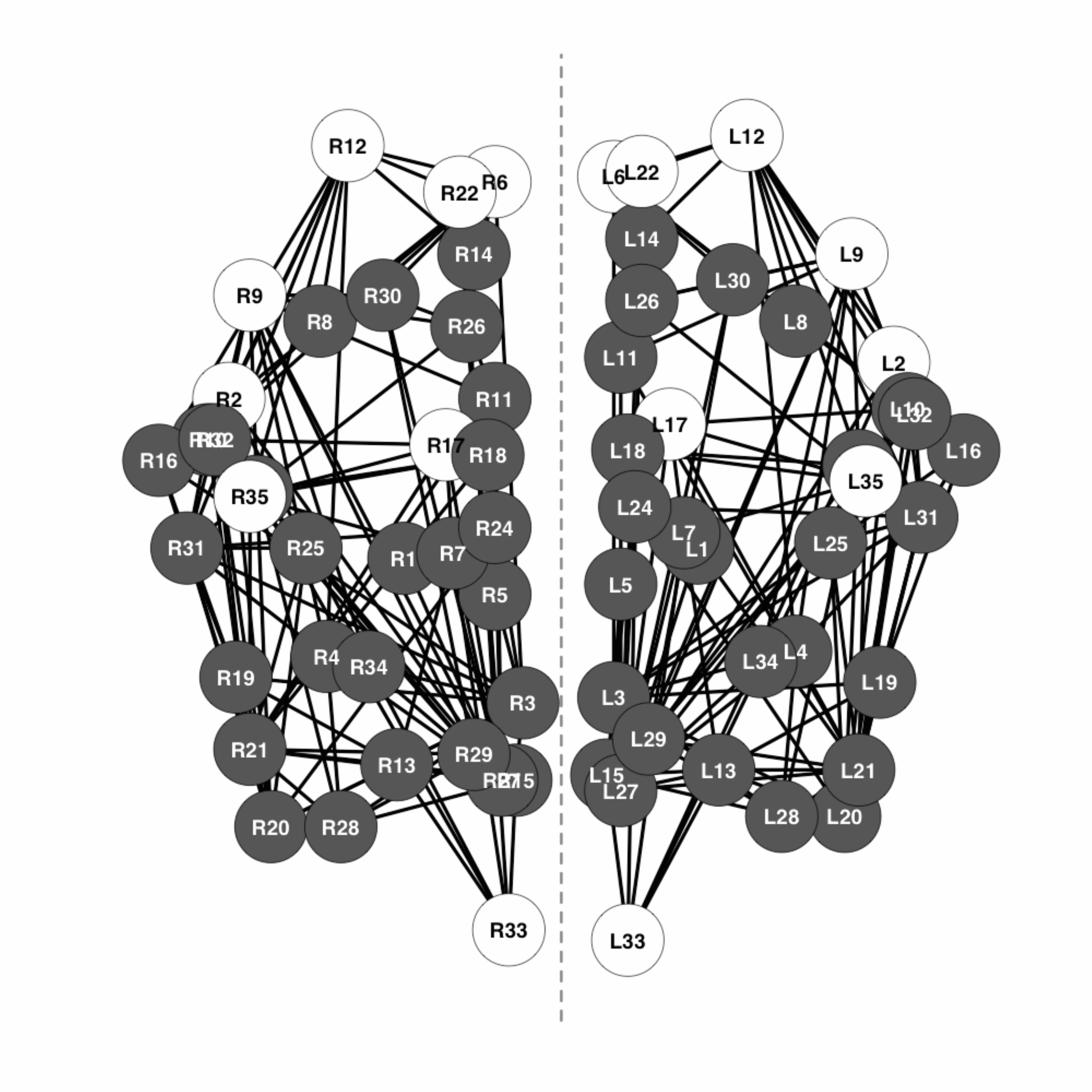}\includegraphics[width=0.46\textwidth]{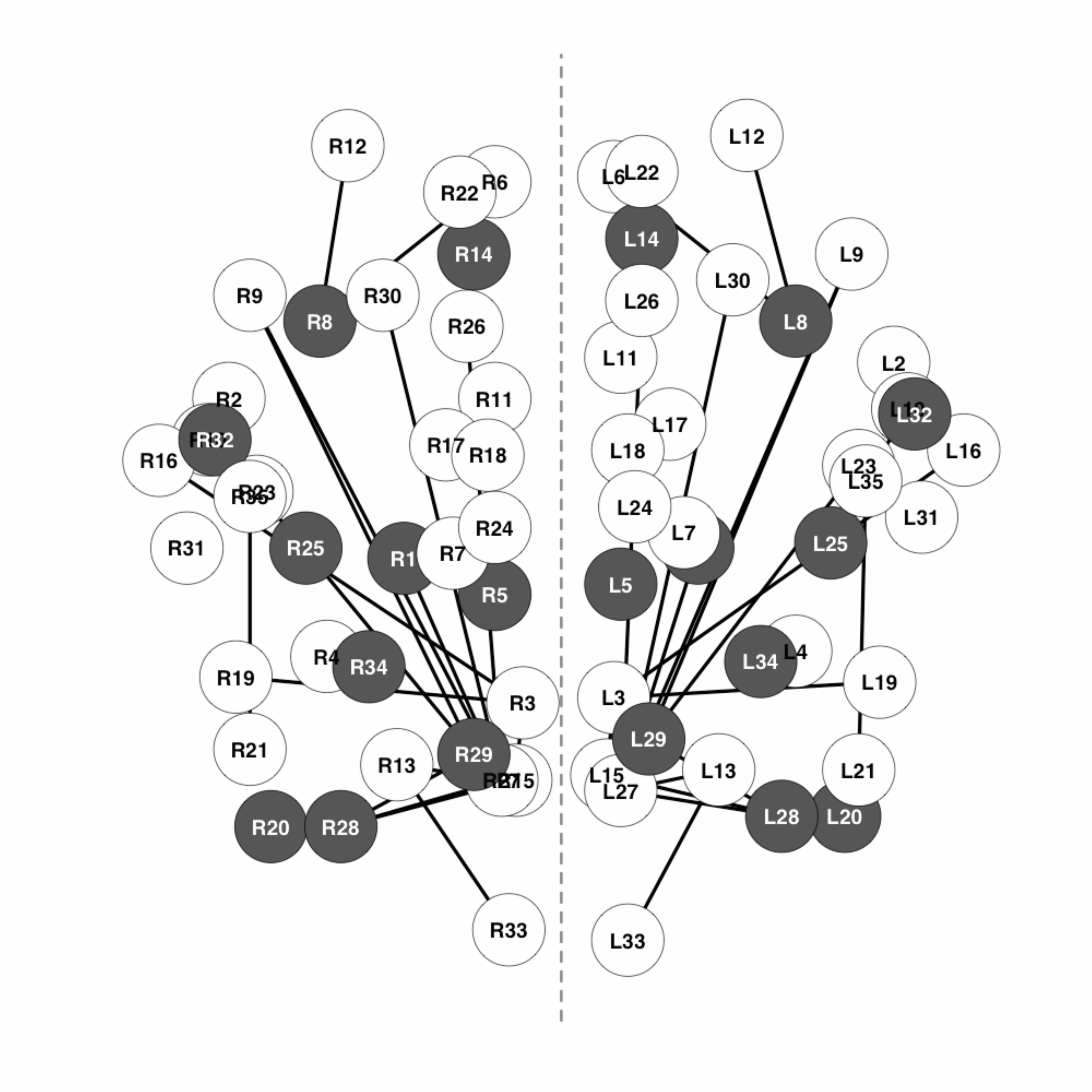}
\\
\includegraphics[width=0.46\textwidth]{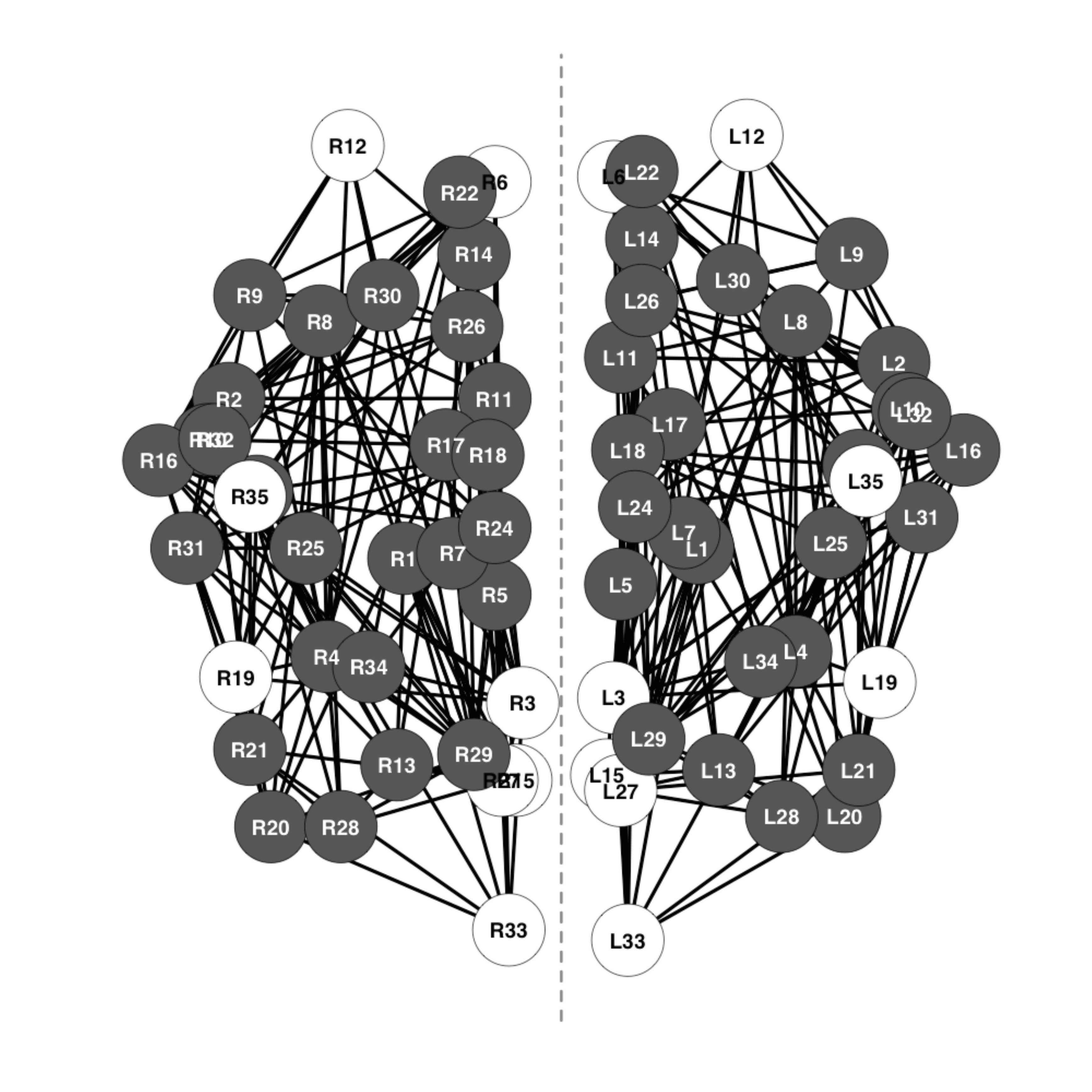}\includegraphics[width=0.46\textwidth]{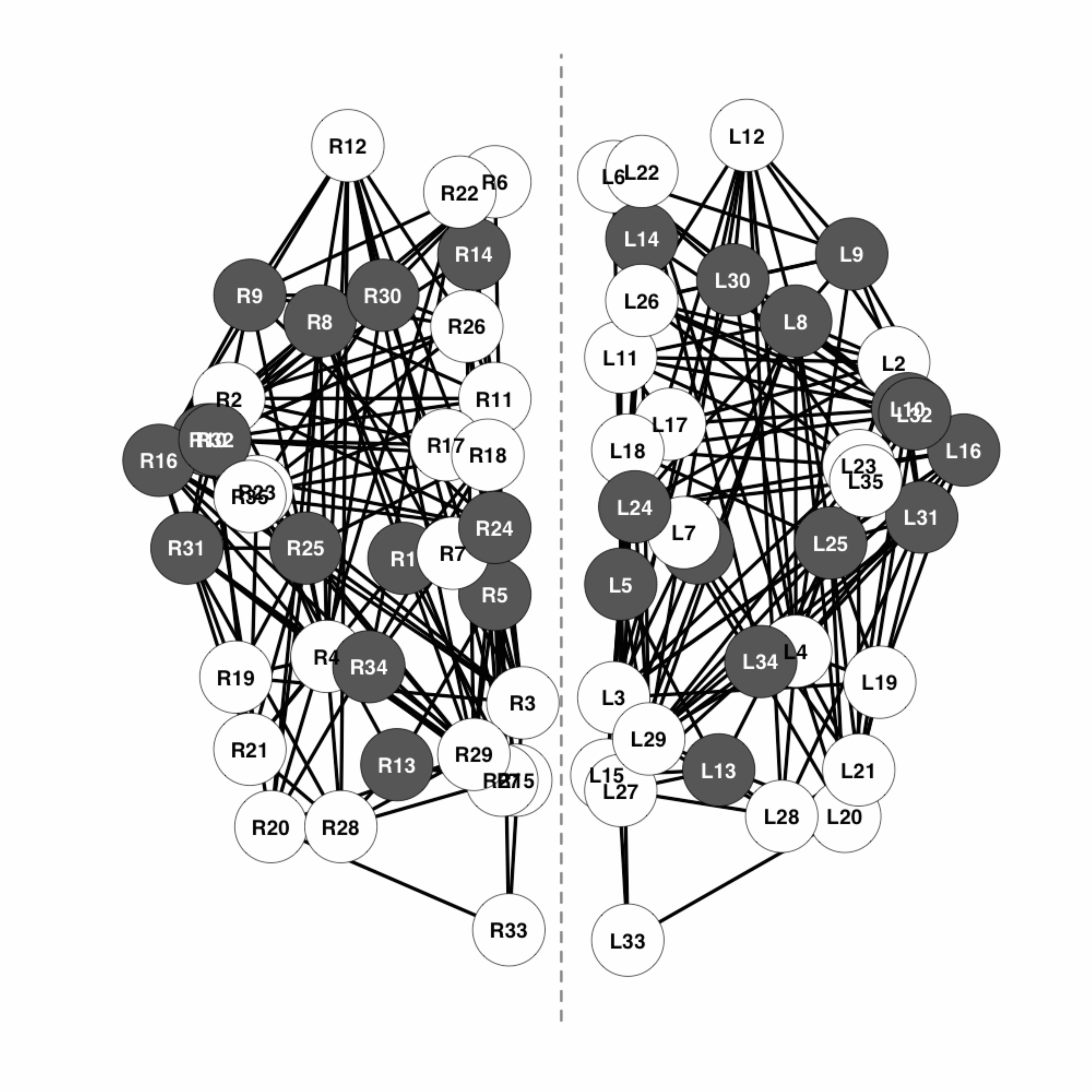}\includegraphics[width=0.46\textwidth]{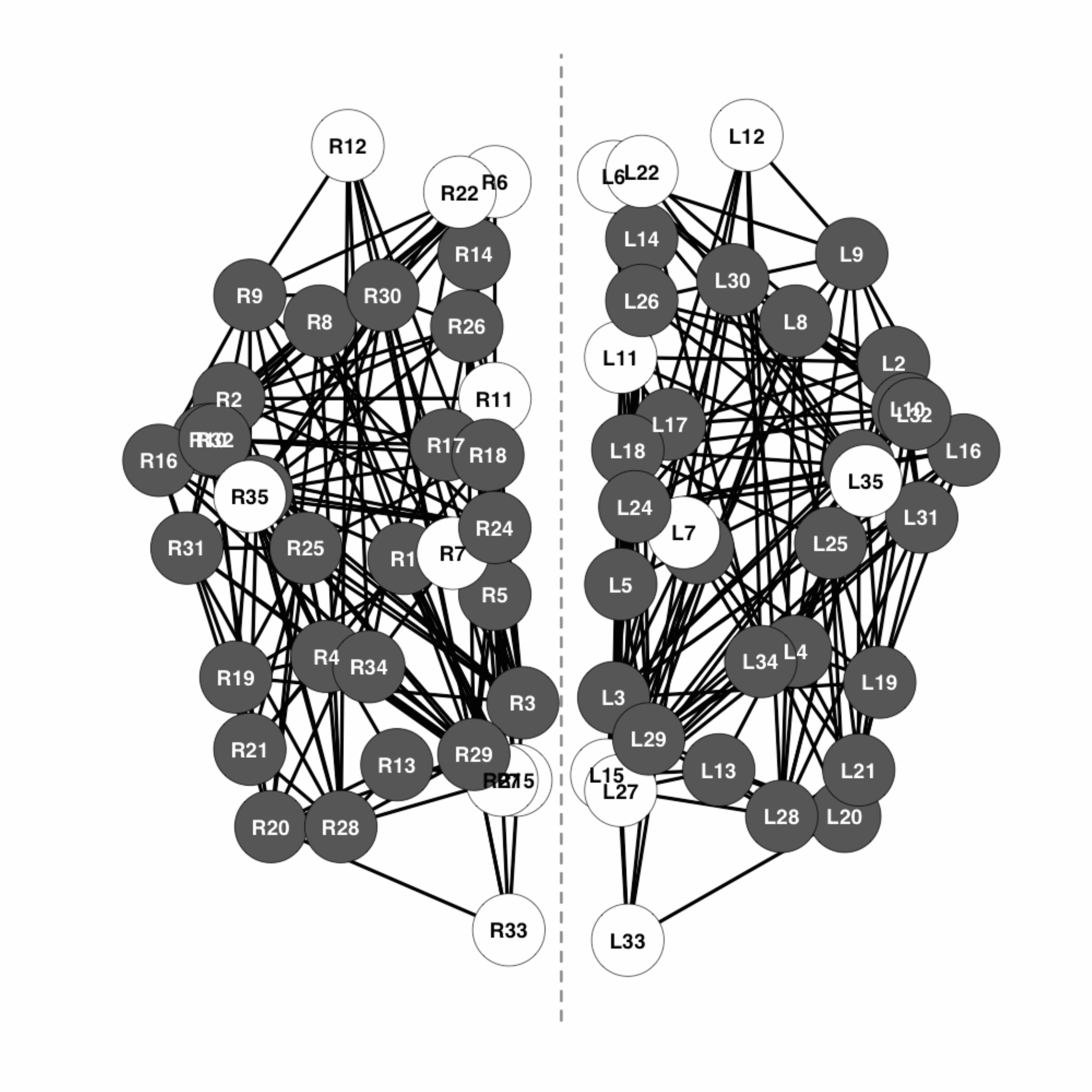}
\caption{Graphical representation of the models in Tables~\ref{tab_var}, \ref{tab_dsc} and \ref{tab_lpr} obtained from the application of symmetric graphical lasso with eBIC. Edges encode symmetric nonzero concentrations whereas shaded vertices represent symmetric diagonal concentrations. From left to right: VAR(1); score driven model; local polynomial regression - Henderson filter with $p=3$ and $h=6$. From top to bottom: subject 18; subject 22.}
\label{fig_tot}
\end{figure}
\end{landscape}

\begin{figure}
\centering
\includegraphics[width=0.95\textwidth]{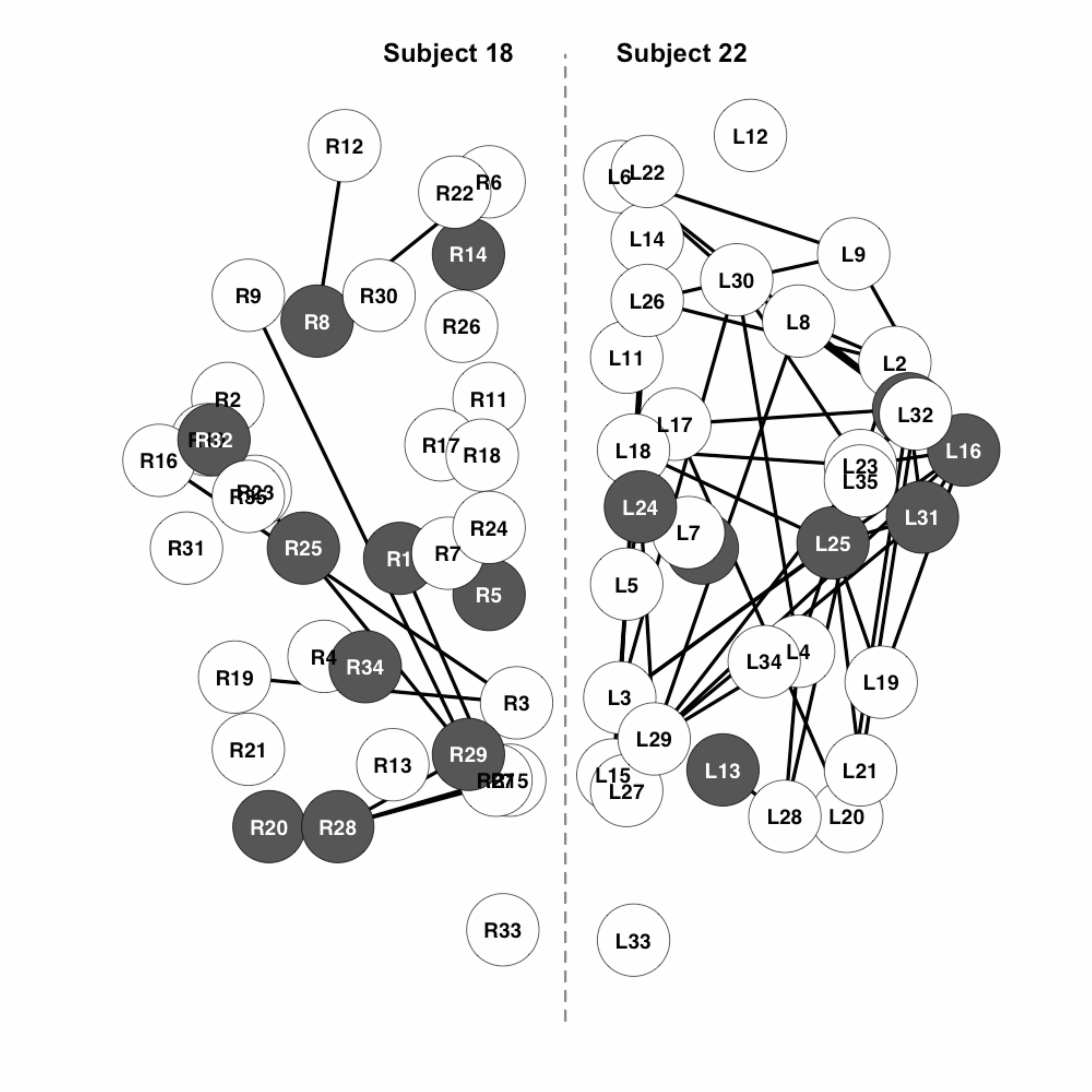}
\caption{\label{fig_commons2}  Common graph structure across the three different filter techniques obtained from the intersection of the graphs in Figure~\ref{fig_tot}, for subject 18 (\emph{left of vertical dashed line}) and subject 22 (\emph{right of vertical dashed line}).}
\end{figure}

The empirical results of graphical lasso (\texttt{gl}) and symmetric graphical lasso (\texttt{sgl}) fit on the residuals estimates for the two subjects are reported in Table \ref{tab_var} (vector autoregressive model, VAR), Table \ref{tab_dsc} (score driven model, DCS) and Table  \ref{tab_lpr} (Henderson filter, H13), according to the different filtering techniques.  For sake of comparison, we report the results obtained by using both eBIC and its limit value BIC. However, we focus on the models obtained from the minimization of eBIC that are more parsimonious than the corresponding models selected from BIC, in particular for the symmetric graphical lasso.
A more direct visual representation of the results detailed in Tables \ref{tab_var}, \ref{tab_dsc} and \ref{tab_lpr} is summarised in Figures \ref{fig_tot} and \ref{fig_commons2}, which provide a graphical representation of the brain symmetry structure. Specifically, the edges of the graphs encode symmetric nonzero off-diagonal concentrations whereas shaded vertices denote symmetric diagonal concentrations. While Figure \ref{fig_tot} summarises the results for the two subjects across the three filtering methods, Figure \ref{fig_commons2} reports the symmetries which turn out to be common to the three methods for subject 18 (left) and 22 (right). Moreover, in every Figure we omit non symmetric edges from the visualization, in order to highlight the novelty aspect of the analysis and also to facilitate the reader with graphs that would otherwise be too densely plotted to be appreciable; nevertheless, the overall edge density for each result are reported in the summary Tables \ref{tab_var} to \ref{tab_lpr}.

The first evident result is that, regardless of the filtering method, subject 22 shows a denser and more symmetric graph than subject 18. Also, the three models selected by eBIC for subject 22 have both similar densities and similar amount of symmetric edges and nonzero concentrations; compare, for instance, the 649 edges of subject 22 and the 373 edges of subject 18 in Table \ref{tab_var},  related to VAR estimation. On the other hand, filtering has an impact on subject 18, as it is evident that the three methods induce a different density (from 11.68\% of H13 in Table \ref{tab_lpr}, to 21.70\% of DCS  in Table \ref{tab_dsc}) and a different amount of symmetries in the resulting graph (from the 42 pairs of symmetric edges of H13  to the 123 of DCS).
These considerations are ever more
pronounced in Figure \ref{fig_tot}, which shows how the graph associated with local polynomial regression is the one exhibiting the lowest number of symmetric concentrations, both diagonal (shaded nodes) and off-diagonal (black solid lines). The second lowest number of symmetric concentrations is observed for the graph identified from the residuals of a VAR(1) model, whereas the residuals from the score driven model bring to a graph with the highest number of symmetric off-diagonal concentrations. This is in line with the idea that a robust detrending method leaves more information in the residuals; on the other hand, methods that tend to overfit the data, such as an high-degree local polynomial regression, may allocate most of the data dependence structure in the signal component, rather than in the noise. This can be quantitatively assessed by comparing the number of pairs of symmetrical concentrations, both off-diagonal and diagonal, reported as the last two columns of Tables~\ref{tab_var}, \ref{tab_dsc} and \ref{tab_lpr}.

Using different filtering techniques also allows one to extract the common information on the symmetric structure across the three filtering methods, by retaining the shared graph of the concentrations for each subject, i.e. the graph resulting from the intersection of the three graphs in Figure~\ref{fig_tot}. After this marginalization, and in order to give a better insight, we
juxtapose them in Figure~\ref{fig_commons2}.
In this side-by-side comparison, subject 22 exhibits an higher number of off-diagonal concentrations,
and it is worth noting both subject seems to share approximately the same number of core
symmetric diagonal values.

In conclusion, we may envisage two main empirical findings emerging from the analysis. The first evidence is concerned with the fact that the subject with a diagnosis of disorder shows a more symmetric brain structure than the healthy one. This is in line with several studies that are in favour of an evidence of lack of asymmetry in schizophrenic patients, see  \cite{su2015}. Nevertheless, the literature is quite controversial on this theme, see the review paper by \citet[][Section~3.1.4]{stephane2001} and our results just refer to a pair of subjects. Secondly,
the impact of the detrending method appears to be stronger in a subject who presents a less symmetric brain structure while it seems to be irrelevant in the subject who shows a more defined symmetric pattern.
In any case, the combination of different filtering methods may shed light on the core symmetries that characterize the brain of different subjects.

\section{Discussion}\label{sec:discussion}
The procedures introduced in this paper have been developed with a focus on the identification of brain networks form RfMRI data. In this respect, we have first considered the problem of removing the temporal dependence from data and then
we have designed our methods and algorithms to suit the natural partition of the brain into hemispheres.
Nevertheless, symmetric graphical lasso identifies a model within the class of RCON models and, as such, it has a potentially wider range of applications. To the best of our knowledge, the symmetric graphical lasso proposed in this contribution is the first instance where the lasso procedure is used to perform model selection within the class of colored graphical models. In this way, we are able to identify symmetries characterized by equality constraints in the entries of the concentration matrix. Methods not explicitly designed to identify symmetric nonzero concentrations, such as the graphical lasso, can still identify symmetries in the graph structure but only in terms of edges being present or absent. Tables \ref{tab_var}-\ref{tab_dsc}-\ref{tab_lpr} compare the graphical lasso and the symmetric graphical lasso in their ability to identify symmetric edges, and show that our method encourages structural symmetries without affecting the graph sparsity.

Our proposed approach has an associated computational complexity comparable with that of methods using penalties of similar type, such as a conventional fused or group lasso \citep{danaher2014joint}. The computational effort associated to using the ADMM algorithm to solve this class of convex quadratic programming optimizations is admittedly the eigen decomposition of a $p \times p$ matrix \citep{tibshirani2005sparsity}. In the application considered in this manuscript, the dimensionality of the problem is bounded by the atlases used in resting state fMRI studies: usually, these atlases identify a number of ROIs close to $p=70,$ as in the data analyzed in Section 5, which makes estimation times of our procedure not a concern. If indeed other applications are considered, such as cancer genomic where the number of variables is in the order of thousands, then exploring some potential pre-screening procedures would be a very interesting avenue to pursue, maybe adapting those mentioned in \cite{danaher2014joint} and \cite{yang2015fused} to our symmetric penalty.

Future research directions involve the specification of a convex penalty function that allows a more flexible specification of color classes, which could affect other sub-components of the main concentration matrix, as well as consider different types of constraints.

\section*{Software}
The code implementing the ADMM algorithm described in this paper is available at the following \texttt{GitHub} repository: \url{https://github.com/savranciati/sgl}.

\section*{Acknowledgements}
The authors would like to thank S\o ren H\o jsgaard for the support provided with \texttt{gRc} package, Veronica Vinciotti for useful discussions, as well as both referees and associated editor for providing feedbacks that improved the quality of the paper. Authors AR and SR were supported by the Air Force Office of Scientific Research under award number FA9550-17-1-0039.
We would also like to thank Antonio Canale, Daniele Durante, Lucia Paci and Bruno Scarpa for  introducing us to the challenging dataset analysed in the paper. These data were provided by Greg Kiar and Eric Bridgeford from NeuroData at Johns Hopkins University, who graciously pre-processed the raw DTI and R-fMRI imaging data available at
\url{http://fcon_1000.projects.nitrc.org/indi/CoRR/html/nki_1.html}, using the pipelines ndmg and C-PAC.

\bibliographystyle{rss}
\bibliography{main_biblio_saverio,main_biblio_alberto,main_biblio_ale}

\newpage

\begin{table}
\caption{\label{tab_sim1} Performance measures in (\ref{eqn:four.performance.measures-E}) and (\ref{eqn:four.performance.measures-S}) from the application of graphical lasso (\texttt{gl}) and symmetric graphical lasso (\texttt{sgl}) to one of the datasets in the simulated scenario A. All values, except \#edges and \#symm, are reported in percentages and the line in bold corresponds to the model for which sTPR+sTNR is maximal.}
\centering
\begin{tabular}{ccccc|ccccc}
\hline
\multirow{2}{*}{Method} & \multicolumn{4}{c|}{Graph structure} & \multicolumn{5}{c}{Symmetric nonzero concentrations} \\
\cline{2-10}
&  ePPV & eTPR & eTNR &  \#edges & sPPV & sTPR & sTNR & sTPR+sTNR & \#symm \\
\hline
\texttt{gl} & 56.09 & 56.29 & 86.82 &  558 & - & - & - & - & - \\
\hline
 & 56.01 & 56.12 & 86.82 &  557 & 0.00 & 0.00 & 100.00 & 100.00 & 0 \\
 & 55.94  & 55.94 & 86.82 &  556 & 0.00 & 0.00 & 100.00 & 100.00 & 0 \\
\multirow{7}{*}{
\begin{minipage}[t]{0.12\columnwidth}
\centering {\texttt{sgl}} \\
{\footnotesize \emph{(increasing \\ values of $\lambda_2$)}}
\vspace{2.5cm}
\end{minipage}}
 & 56.39 & 55.58 & 87.14 &  548 & 0.00  & 0.00 & 99.65 & 99.65 & 2 \\
 & 57.89 & 55.40 & 87.95 &  532 & 25.00 & 1.67 & 98.77 & 100.44 & 8 \\
& 58.98 & 54.32 & 88.70 &  512 & 41.67 & 8.33 & 97.37 & 105.70 & 20 \\
 & 60.16 & 54.86 & 89.13 &  507 & 42.86 & 20.00 & 95.26 & 115.26 & 39 \\
 & \textbf{56.45} & \textbf{50.36} & \textbf{88.38} &  \textbf{496} & \textbf{37.31} & \textbf{41.67} & \textbf{89.82} & \textbf{131.49} & \textbf{83} \\
 & 53.51 & 48.02 & 87.52 &  499 & 28.87 & 46.67 & 82.28 & 128.95 & 129 \\
 & 52.98 & 48.02 & 87.25 &  504 & 28.00 & 46.67 & 81.23& 127.90 & 135 \\
 & 52.98 & 48.02  & 87.25 &  504 & 28.00 & 46.67 & 81.23& 127.90 & 135 \\
\hline
\end{tabular}
\end{table}

\begin{landscape}
\begin{table}
\caption{\label{tab_sim2} Performance measures in (\ref{eqn:four.performance.measures-E}) and (\ref{eqn:four.performance.measures-S}) from the application of graphical lasso (\texttt{gl}) and symmetric graphical lasso (\texttt{sgl}) to the $8\times 9$ simulated datasets. Data are reported as mean (and standard deviation) computed across the 9 replicated datasets for each of the 8 environments. Values are reported in percentages for ePPV, eTPR, eTNR, and density; average number of edges, average number of symmetric concentrations, and their standard deviations are rounded.}
\centering
\begin{tabular}{cccccccc|cccc}
\hline
\multicolumn{2}{c}{Environment} & \multirow{2}{*}{Method} & \multicolumn{5}{c|}{Graph structure} & \multicolumn{4}{c}{Symmetric nonzero concentrations} \\
\cline{1-2}\cline{4-12}
Scenario & Code &  & ePPV & eTPR & eTNR & \#edges & Density & sPPV & sTPR & sTNR & \#symm \\
\hline
\multirow{8}{*}{A} & \multirow{2}{*}{A.1} & \texttt{gl} & 57.8 {\footnotesize (1.3)} & 57.9 {\footnotesize (1.3)} &  87.3 {\footnotesize (0.4)} & 560 {\footnotesize \,\,(1)} & 23.2 {\footnotesize (0.1)} & - & - & - & -  \\
& & \texttt{sgl} & 54.8 {\footnotesize (1.9)} & 47.4 {\footnotesize (1.7)} & 88.24 {\footnotesize (0.6)} & 483 {\footnotesize \,\,(9)} & 20.0 {\footnotesize (0.4)} & 30.6 {\footnotesize (4.8)}  & 49.5 {\footnotesize (5.7)}  & 87.9 {\footnotesize (2.7)} & 99 {\footnotesize (18)}  \\
\cline{2-12}
& \multirow{2}{*}{A.2} & \texttt{gl} & 57.6 {\footnotesize (1.1)} & 57.6 {\footnotesize (1.2)} &  87.6 {\footnotesize (0.3)} & 547 {\footnotesize \,\,(1)} & 22.7 {\footnotesize (0.1)} & - & - & - & -  \\
& & \texttt{sgl} & 54.7 {\footnotesize (2.1)} & 48.2 {\footnotesize (2.4)} & 88.3 {\footnotesize (0.6)} & 482 {\footnotesize (12)} & 20. {\footnotesize (0.5)} & 28.1 {\footnotesize (5.9)}  & 44.4 {\footnotesize (4.4)}  & 87.4 {\footnotesize (3.6)} & 98 {\footnotesize (21)}  \\
\cline{2-12}
&\multirow{2}{*}{A.3} & \texttt{gl} & 57.6 {\footnotesize (1.5)} & 57.6 {\footnotesize (1.4)} &  87.3 {\footnotesize (0.5)} & 556 {\footnotesize \,\,(1)} & 23.0 {\footnotesize (0.1)} & - & - & - & -  \\
& & \texttt{sgl} & 55.2 {\footnotesize (2.5)} & 47.46 {\footnotesize (2.4)} & 88.4 {\footnotesize (1.0)} & 479 {\footnotesize (23)} & 19.8 {\footnotesize (1.0)} & 25.5 {\footnotesize (4.2)}  & 41.3 {\footnotesize (6.2)}  & 86.8 {\footnotesize (3.3)} & 100 {\footnotesize (21)}  \\
\cline{2-12}
& \multirow{2}{*}{A.4} & \texttt{gl} & 55.8 {\footnotesize (1.2)} & 57.8 {\footnotesize (1.3)} &  86.7 {\footnotesize (0.4)} & 559 {\footnotesize \,\,(1)} & 23.1 {\footnotesize (0.1)} & - & - & - & -  \\
& & \texttt{sgl} & 54.0 {\footnotesize (1.3)} & 47.7 {\footnotesize (1.4)} & 87.8 {\footnotesize (0.6)} & 494 {\footnotesize (14)} & 20.5 {\footnotesize (0.6)} & 29.5 {\footnotesize (5.3)}  & 43.9 {\footnotesize (4.5)}  & 88.6 {\footnotesize (2.7)} & 92 {\footnotesize (17)} \\
\hline \hline \hline \hline
\multirow{8}{*}{B} & \multirow{2}{*}{B.1} & \texttt{gl} & 57.4 {\footnotesize (0.6)} & 57.4 {\footnotesize (0.6)} &  80.4 {\footnotesize (0.3)} & 763 {\footnotesize \,\,(2)} & 31.6 {\footnotesize (0.1)} & - & - & - & -  \\
& & \texttt{sgl} & 56.0 {\footnotesize (1.3)} & 52.3 {\footnotesize (1.2)} & 81.1 {\footnotesize (0.9)} & 711 {\footnotesize (19)} & 29.5 {\footnotesize (0.8)} & 54.1 {\footnotesize (5.4)}  & 40.9 {\footnotesize (3.2)}  & 85.9 {\footnotesize (3.8)} & 137 {\footnotesize (22)}  \\
\cline{2-12}
&  \multirow{2}{*}{B.2} & \texttt{gl} & 57.8 {\footnotesize (1.0)} & 57.8 {\footnotesize (1.0)} &  81.1 {\footnotesize (0.5)} & 748 {\footnotesize \,\,(1)} & 31.0 {\footnotesize (0.1)} & - & - & - & -  \\
& & \texttt{sgl} & 56.0 {\footnotesize (1.6)} & 51.9 {\footnotesize (1.8)} & 81.7 {\footnotesize (0.7)} & 694 {\footnotesize (12)} & 28.7 {\footnotesize (0.5)} & 54.4 {\footnotesize (2.4)}  & 40.4 {\footnotesize (3.6)}  & 86.6 {\footnotesize (1.9)} & 133 {\footnotesize (14)}  \\
\cline{2-12}
&  \multirow{2}{*}{B.3} & \texttt{gl} & 58.1 {\footnotesize (0.9)} & 58.2 {\footnotesize (0.9)} &  80.8 {\footnotesize (0.4)} & 758 {\footnotesize \,\,(1)} & 31.4 {\footnotesize (0.1)} & - & - & - & -  \\
& & \texttt{sgl} & 56.8 {\footnotesize (1.1)} & 52.6 {\footnotesize (0.7)} & 81.7 {\footnotesize (0.8)} & 702 {\footnotesize (15)} & 29.1 {\footnotesize (0.6)} & 53.7 {\footnotesize (5.5)}  & 41.4 {\footnotesize (3.3)} & 85.4 {\footnotesize (4.0)} & 140 {\footnotesize (23)}  \\
\cline{2-12}
& \multirow{2}{*}{B.4} & \texttt{gl} & 58.2 {\footnotesize (1.5)} & 58.2 {\footnotesize (1.5)} &  81.3 {\footnotesize (0.7)} & 745 {\footnotesize \,\,(1)} & 30.9 {\footnotesize (0.1)} & - & - & - & -  \\
& & \texttt{sgl} & 57.0 {\footnotesize (1.7)} & 53.2 {\footnotesize (1.9)} & 82.1 {\footnotesize (0.8)} & 695 {\footnotesize (14)} & 28.8 {\footnotesize (0.6)} & 52.7 {\footnotesize (3.7)}  & 44.5 {\footnotesize (4.5)}  & 84.0 {\footnotesize (2.7)} & 152 {\footnotesize (17)}  \\
\hline
\end{tabular}
\end{table}
\end{landscape}

\begin{table}
\caption{\label{tab_var} Empirical results for graphical lasso (\texttt{gl}) symmetric graphical lasso (\texttt{sgl}) fit on residuals from a VAR(1) model on two subjects. The last three columns provide a description of the symmetric structure, that is the number of pairs of symmetric: (i) edges, (ii) off-diagonal nonzero concentrations and (iii) diagonal concentrations.}
\centering
\begin{tabular}{ccccc|ccc}
\hline
\multirow{3}{*}{Subject} &  \multirow{3}{*}{Criterion} &  \multirow{3}{*}{Method} & \multirow{3}{*}{\#edges} & \multirow{3}{*}{Density} & \multicolumn{3}{c}{Pairs of symmetric} \\
\cline{6-8}
& & & & & \multirow{2}{*}{edges} & \multicolumn{2}{c}{nonzero concentrations}\\
\cline{7-8}
& & & & & & off-diagonal & diagonal \\
\hline
\multirow{4}{*}{18} & \multirow{2}{*}{BIC} & \texttt{gl} & 876 & 36.27\% & 120 & - & - \\
&  &  \texttt{sgl} & 910 & 37.68\% & 171 & 87 & 6 \\
\cline{2-8}
&  \multirow{2}{*}{eBIC $\gamma=0.5$}  & \texttt{gl}  & 366 & 15.16\% & 48 & - & - \\
&  &  \texttt{sgl} & 373 & 15.45\% & 92 & 89 & 31 \\
\hline \hline \hline
\multirow{4}{*}{22} & \multirow{2}{*}{BIC}  & \texttt{gl}  & 879 & 36.40\% & 128 & - & - \\
&  &  \texttt{sgl} & 891 & 36.89\% & 202 & 160 & 14 \\
\cline{2-8}
&  \multirow{2}{*}{eBIC $\gamma=0.5$} & \texttt{gl} & 624 & 25.84\% & 91 & - & - \\
&  &  \texttt{sgl} & 649 & 26.87\% & 158 & 149 & 27 \\
\hline
\end{tabular}
\end{table}

\begin{table}
\caption{\label{tab_dsc} Empirical results for graphical lasso (\texttt{gl}) symmetric graphical lasso (\texttt{sgl}) fit on residuals from a score driven model on two subjects. The last three columns provide a description of the symmetric structure, that is the number of pairs of symmetric: (i) edges, (ii) off-diagonal nonzero concentrations and (iii) diagonal concentrations.}
\centering
\begin{tabular}{ccccc|ccc}
\hline
\multirow{3}{*}{Subject} &  \multirow{3}{*}{Criterion} &  \multirow{3}{*}{Method} & \multirow{3}{*}{\#edges} & \multirow{3}{*}{Density} & \multicolumn{3}{c}{Pairs of symmetric} \\
\cline{6-8}
& & & & & \multirow{2}{*}{edges} & \multicolumn{2}{c}{nonzero concentrations}\\
\cline{7-8}
& & & & & & off-diagonal & diagonal \\
\hline
\multirow{4}{*}{18} &\multirow{2}{*}{BIC} & \texttt{gl} & 815 & 33.75\% & 115 & - & - \\
&  &  \texttt{sgl} & 826 & 34.20\% & 195 & 175 & 22 \\
\cline{2-8}
&  \multirow{2}{*}{eBIC $\gamma=0.5$}  & \texttt{gl} & 513 & 21.24\% & 72 & - & - \\
&  &  \texttt{sgl} & 524 & 21.70\% & 123 & 120 & 27 \\
\hline \hline \hline
\multirow{4}{*}{22} & \multirow{2}{*}{BIC}  & \texttt{gl}  & 894 & 37.02\% & 121 & - & - \\
&  &  \texttt{sgl} & 885 & 36.65\% & 167 & 110 & 8 \\
\cline{2-8}
&  \multirow{2}{*}{eBIC $\gamma=0.5$}  & \texttt{gl} & 625 & 25.88\% & 91 & - & - \\
&  &  \texttt{sgl} & 640 & 26.50\% & 149 & 134 & 14 \\
\hline
\end{tabular}
\end{table}

\begin{table}
\caption{\label{tab_lpr} Empirical results for graphical lasso (\texttt{gl}) symmetric graphical lasso (\texttt{sgl}) fit on residuals from a local polynomial regression with $p=3$ - Henderson filter weights - and $h=6$ on two subjects. The last three columns provide a description of the symmetric structure, that is the number of pairs of symmetric: (i) edges, (ii) off-diagonal nonzero concentrations and (iii) diagonal concentrations.}
\centering
\begin{tabular}{ccccc|ccc}
\hline
\multirow{3}{*}{Subject} &  \multirow{3}{*}{Criterion} &  \multirow{3}{*}{Method} & \multirow{3}{*}{\#edges} & \multirow{3}{*}{Density} & \multicolumn{3}{c}{Pairs of symmetric} \\
\cline{6-8}
& & & & & \multirow{2}{*}{edges} & \multicolumn{2}{c}{concentrations}\\
\cline{7-8}
& & & & & & off-diagonal & diagonal \\
\hline
\multirow{4}{*}{18} & \multirow{2}{*}{BIC} & \texttt{gl} & 940 & 38.82\% & 133 & - & - \\
& &  \texttt{sgl} & 937 & 38.80\% & 188 & 135 & 15 \\
\cline{2-8}
& \multirow{2}{*}{eBIC $\gamma=0.5$}  & \texttt{gl} & 285 & 11.80\% & 33 & - & - \\
& &  \texttt{sgl} & 282 & 11.68\% & 42 & 21 & 10 \\
\hline \hline \hline
\multirow{4}{*}{22} &\multirow{2}{*}{BIC} & \texttt{gl} & 913 & 37.81\% & 119 & - & - \\
&  &  \texttt{sgl} & 870 & 36.02\% & 173 & 123 & 10 \\
\cline{2-8}
&  \multirow{2}{*}{eBIC $\gamma=0.5$}  & \texttt{gl} & 644 & 26.67\% & 81 & - & - \\
& &  \texttt{sgl} & 645 & 26.71\% & 156 & 145 & 26 \\
\hline
\end{tabular}
\end{table}

\end{document}